\documentclass[useAMS,usenatbib]{mn2e}
\usepackage{graphicx,color,epsfig}

\newcommand{\be}{\begin{equation}}
\newcommand{\ee}{\end{equation}}

\newcommand{\apj}{ApJ}
\newcommand{\apjs}{ApJS}
\newcommand{\mnras}{MNRAS}
\newcommand{\aap}{A\&A}
\newcommand{\araa}{ARA\&A}
\newcommand{\apjl}{ApJL}

\newcommand{\nat}{Nature}

\newcommand{\icarus}{ICARUS}

\def\ltsima{$\; \buildrel < \over \sim \;$}
\def\simlt{\lower.5ex\hbox{\ltsima}}
\def\gtsima{$\; \buildrel > \over \sim \;$}
\def\simgt{\lower.5ex\hbox{\gtsima}}

\newcommand\mearth{{\,{\rm M}_{\oplus}}}
\newcommand\mj{{\,{\rm M}_{\rm J}}}

\def\msun{{\,M_\odot}}

\title[Tidal Downsizing III]{Tidal Downsizing Model. III. Planets from
  sub-Earths to Brown Dwarfs: structure and metallicity preferences}

\author[Nayakshin \& Fletcher]{Sergei Nayakshin and Mark Fletcher\\ Department of Physics \& Astronomy,
  University of Leicester, Leicester, LE1 7RH, UK\\ {E-mail:~} {\rm
    Sergei.Nayakshin@le.ac.uk}}

\begin{document}

\date{Received}

\pagerange{\pageref{firstpage}--\pageref{lastpage}} \pubyear{2008}

\maketitle

\label{firstpage}

\begin{abstract}
We present population synthesis calculations of the Tidal Downsizing (TD)
hypothesis for planet formation.  Our models address the following
observations: (i) most abundant planets being Super Earths; (ii) cores more
massive than $\sim 5-15 M_\oplus$ are enveloped by massive atmospheres; (iii)
the frequency of occurrence of close-in gas giant planets correlates strongly
with metallicity of the host star; (iv) no such correlation is found for
sub-Neptune planets; (v) presence of massive cores in giant planets; (vi) gas
giant planets are over-abundant in metals compared to their host stars; (vii)
this over-abundance decreases with planet's mass; (viii) a deep valley in the
planet mass function between masses of $\sim 10-20 M_\oplus$ and $\sim 100
M_\oplus$.  A number of observational predictions distinguish the model from
Core Accretion: (a) composition of the massive cores is always dominated by
rocks not ices; (b) the core mass function is smooth with no minimum at $\sim
3 M_\oplus$ and has no ice-dominated cores; (c) gas giants beyond 10 AU are
insensitive to the host star metallicity; (d) objects more massive than $\sim
10 M_{\rm Jup}$ do not correlate or even anti-correlate with metallicity. The
latter prediction is consistent with observations of low mass stellar
companions. TD can also explain formation of planets in close binary systems.
TD model is a viable alternative to the Core Accretion scenario in explaining
many features of the observed population of exoplanets.
\end{abstract}


\section{Introduction}\label{sec:intro}

Core Accretion model \citep[CA; e.g.,][]{PollackEtal96,AlibertEtal05}
stipulates that all planets grow from planetesimals -- rocky or icy bodies
$\sim 1$~km or more in size \citep{Safronov72}. Planetesimals combine into
bigger solid bodies by sticking collisions. More recent work suggests that
planetesimals may have formed via streaming instabilities
\citep{YoudinGoodman05,JohansenEtal07} and were born big, e.g., as large as
$\sim 100$ to 1000 km in size \citep{MorbidelliEtal09}.  In addition to this,
pebbles, which are grains that have grown to the size of $\sim$ 1 mm to a few
cm, are now suspected to contribute to the growth of the cores strongly
\citep{OrmelKlahr10,LambrechtsJ12,Chambers14,LambrechtsEtal14}.

Whatever the growth mechanism of the cores, those that become massive
attract gaseous atmospheres from protoplanetary discs. The atmosphere
  eventually becomes as massive as the core when the core mass exceeds a
  critical value, $M_{\rm crit} \sim 10\mearth$, although the exact critical
  core mass is a function of dust opacity, planet's location and other
  important physics
  \citep[e.g.,][]{PerriCameron74,Mizuno80,Stevenson82,Rafikov06}. At this
  point a runaway accretion of gas onto the core takes place, forming a gas
  giant planet \citep{PollackEtal96,HubickyjEtal05}.

CA is the most widely accepted theory of planet formation \citep[e.g.,
  see][]{HelledEtal13a}. CA popularity is motivated by its successes and, in no
small measure, by the failures of the alternative model, Gravitational disc
Instability \citep[GI; e.g.,][]{Kuiper51,CameronEtal82,Boss97}. In particular,
the classical version of GI cannot account for (i) the existence of
terrestrial/rocky planets; (ii) any planets within the inner $\sim$ tens of AU
of the host star; (iii) presence of massive cores inside of and metal
overabundance of gas giant planets, and (iv) a positive giant planet frequency
of occurrence -- host star metallicity correlation
\citep{FischerValenti05}. CA, in contrast, has features (i-iii) built in by
construction and predicts (iv) naturally as a result of producing more massive
cores in high metallicity environments, so that the runaway gas accretion
phase commences earlier \citep{IdaLin04a,IdaLin04b,MordasiniEtal14}.

However, a new planet formation framework called Tidal Downsizing (TD), in
which GI is only the first step, has been suggested relatively recently
\citep{BoleyEtal10,Nayakshin10c}.  The theory is an offspring of the
Gravitational disc Instability model for planet formation
\citep[e.g.,][]{CameronEtal82,Boss97} but is far richer in terms of physics
included in it. In a way, TD theory is GI theory modernised by physical
processes many of which became standard features of CA at various times, but
were somehow forgotten to be included in GI. These processes are planet
migration, solids coagulating into massive bodies, pebble accretion, and
fragment disruption when compromised by too strong tidal forces from the host
star. For a recent review of issues surrounding TD and GI, see
\cite{HelledEtal13a}, and section 2 in \cite{Nayakshin15c}, the latter
specifically focused on TD.

Several ways of addressing problems (i-iii) in the context of TD were
qualitatively clear from its inception, but it is only very recently that a
process accounting for (iv) was found.  We have recently developed detailed
population synthesis models of the new scenario enabling quantitative
comparisons to observations. This paper, third in a series, presents a number
of such comparisons and makes observational predictions that may discriminate
between CA and TD in the future.

The origins of this new theory lie in the apparently forgotten suggestion of
\cite{Kuiper51} that GI may form not only the gas giant planets of the Solar
System but also the rocky ones. He suggested that the inner Solar System
planets are made by destroying $\sim$ Jupiter mass gas fragments within which
dust sedimented down \citep{McCreaWilliams65,Boss98} to form massive and dense
cores composed of heavy elements.  Hydrogen/helium and other volatile
components of the fragments are disrupted by the Solar tides and eventually
consumed by the Sun, whereas the much denser cores survive to become the
present day planets.

Until \cite{BoleyEtal10}, no physical way of actually placing the massive gas
fragments at $\sim$ a few AU distances from the Sun seemed to exist
\citep[e.g.,][]{Rice05,Rafikov05}. This appears to be the main reason why this
avenue of planet formation was discounted early on \citep{DW75}. However, we
now know that the fragments do not have to be born where the planets are now
because of planet migration
\citep[e.g.,][]{LinPap79,GoldreichTremaine80}. \cite{BoleyEtal10} found that
$\sim$ Jupiter mass gas fragments born by gravitational instability in the
outer ($R\sim 100$~AU) cold protoplanetary disc do not stay there, as usually
assumed, but migrate inward rapidly \citep[in as little as $\sim 10^4$ yrs;
  see also][]{BaruteauEtal11,ChaNayakshin11a,ZhuEtal12a}. The fragments are
initially very fluffy, and it takes up to a few Myrs
\citep[e.g.,][]{Bodenheimer74,BodenheimerEtal80,VazanHelled12} for them to
contract and collapse to what is usually taken as the time $t=0$ configuration
of GI planets \citep[the so called "hot start" of gas giants, see,
  e.g.,][]{MarleyEtal07}. Since migration process is much quicker, the
pre-collapse gas fragments are usually disrupted by tides from their host
stars when the fragment separation shrinks to a few AU. \cite{Nayakshin10c}
arrived at similar ideas via analytical estimates and simulations of core
formation within isolated gas fragments \citep{Nayakshin10a,Nayakshin10b}.

Since then, several dozen investigations into the physics of TD or into
processes important to TD \citep[cf. further references in
][]{HelledEtal13a,Nayakshin15c} sprung up. There was however much less work on
trying to relate the TD hypothesis to the observations of exoplanets in a
quantitative way, which is not particularly surprising given the theory is
still in its infancy. Nevertheless, \cite{ForganRice13b} and
\cite{GalvagniMayer14} built the first population synthesis models for TD in
order to compare the predictions of the models with the statistics of observed
exoplanets \citep[for the latter, see a recent review by][]{WF14}. The results
of these two studies are somewhat divergent. \cite{ForganRice13b} find that TD
is incapable of producing any planets at small separations ($R\simlt 10$~AU)
since most of their gas fragments are tidally disrupted {\em before} massive
solid cores could form within them.  \cite{GalvagniMayer14} did not include
grain sedimentation in their models, hence could not say anything about the
post-disruption core-dominated planets. However, their study found a wealth of
Jovian mass planets that successfully migrated into the inner disc, avoiding
tidal disruptions, and hence they suggested that TD may well be effective in
producing a hot-Jupiter like population of gas giants.

Another challenge to TD is the expectation that higher metallicity fragments
are more likely to be tidally disrupted because of slower radiative cooling
\citep{HB11}. This would imply that low metallicity environments must be more
hospitable to gas giant formation via TD channel. Since a strong positive
giant planet frequency -- metallicity correlation is observed
\citep{Gonzalez99,FischerValenti05}, the general feeling is that TD is
strongly disfavoured by the data.

A solution to these and other shortcomings of TD hypothesis may be "pebble
accretion" \citep{Nayakshin15a}, the process in which large ($\sim 1$ mm to a
few cm) grains from the disc separate out of the gas-dust flow past an
embedded massive body (large planetesimal or a planet) and accrete onto it
\citep[e.g.,][]{JohansenLacerda10,OrmelKlahr10,LambrechtsJ12}. Pebble
accretion is mainly called upon in the Core Accretion model to accelerate the
growth of solid cores at large distances
\citep[e.g.,][]{HB14,Chambers14,LambrechtsEtal14}; no application of pebble
accretion to TD was however done until very recently.  

A 1D radiative hydrodynamics study with grains treated as a second fluid
showed that external pebble deposition in pre-collapse gas fragments
significantly accelerates their contraction and collapse
\citep{Nayakshin15a}. Using a 1D viscous disc evolution code to treat the
disc-planet interactions (angular momentum and heat exchange, pebble
accretion) and co-evolution, we confirmed that pebble accretion allows more
fragments to survive tidal disruptions, and also does so preferentially in
high metallicity environments \citep{Nayakshin15b}. In paper I
\citep{Nayakshin15c}, this numerical scheme was extended to include grain
growth, sedimentation and core formation within the fragments. In paper II
\citep{Nayakshin15d}, a large set of population-synthesis like experiments in
TD setting was performed. Since pebble accretion strongly increases grain and
metal abundance within the gas fragments, massive ($M_{\rm core} \sim 10
\mearth$) cores composed of heavy elements form much quicker than they do
inside ``pristine'' fragments made of the dust/gas composition equal to that
of the host star. For this reason, an abundant massive core presence in the
inner disc was found in paper II, in contrast to the earlier results by
\cite{ForganRice13b}.

Interestingly, while survival of gas giant planets in these numerical
experiments was strongly enhanced at high metallicities, the same was not true
for core dominated planets with mass $\simlt 10\mearth$. These planets were
most abundant around $\sim$ Solar metallicity hosts rather than at higher
metallicities. This result is in line with both radial velocity and transit
observations of exoplanets
\citep[e.g.,][]{SousaEtal08,MayorEtal11,BuchhaveEtal12} which showed that
planets smaller than $\sim$ Neptune in mass or radius do not correlate with
metallicity of the host. Strictly speaking, this contradicts expectations from
CA \citep{IdaLin04a} that massive cores are more abundant at high
metallicities. This is the feature of the model called upon to explain the gas
giant planet positive metallicity correlation with metal abundance of the host
stars.  While more recent work \citep{MordasiniEtal12} shows that CA can
explain the absence of metallicity correlation for Neptune mass or smaller
planets, this result depends sensitively on uncertain details of type I
migration prescription (see section 7.2 in the paper quoted just above).

Here we improve on methods of paper I and II in a number of ways, investigate
the results in greater detail, and present additional model predictions to
distinguish TD from CA observationally. First, we study a much broader range
of the initial fragment masses, e.g., from the minimum of $M_{\rm min0} = 1/3
\mj$ to the maximum of $M_{\rm max0} = 16 \mj$ \citep[][studied fragments in a
  much narrower mass range, $0.5 < M_0/\mj < 2$]{Nayakshin15d}. This is
important as the initial fragment mass cannot be accurately constrained at
this stage: different authors arrive at different answers
\citep[e.g.,][]{BoleyEtal10,ForganRice11}. Secondly, following
\cite{NayakshinEtal14a}, we calculate the structure of the dense gas layers
adjacent to the core growing inside the gas fragment, hence resolving the
gravitational influence of the core on the very centre of the gas fragment.
This step is key to studying Neptune to $\sim$ sub-Saturn mass planets. In the
context of TD, such planets form when the core is surrounded by massive gas
atmosphere {\em bound} to the core. This is physically similar to how massive
gas atmospheres form around cores in the context of CA
\citep[e.g.,][]{Mizuno80,Stevenson82,Rafikov06}, but all the action takes
place in the centre of the massive gas fragment (formed by GI and not due to
the presence of the core) rather than the protoplanetary disc.  When the gas
fragment is disrupted, the bound atmosphere survives together with the core
since it is far denser than the rest of the fragment \citep{NayakshinEtal14a}.
Finally, we now use a more realistic -- better observationally constrained --
approach to the protoplanetary disc dispersal to end our runs.

We find that our population synthesis model reproduces a number of
observational facts/correlations for planetary properties that were previously
suggested to vindicate CA as the only correct theory of planet formation, such
as the metallicity correlations, the presence of cores inside of and the
general overabundance of metals in gas giant planets compared to their host
stars, the existence of a critical core mass above which the core must possess
a massive envelope composed of volatiles and hydrogen/helium. Fortunately,
there turn out to be a number of observational predictions distinguishing our
theory from CA model.

Section 2 of this paper presents a summary of the main assumptions and
numerical methods used here. Readers not interested in computational
  detail of the model may skip section 2 and proceed to \S 3, where a broad
brush overview of the results is presented. Planet mass function (PMF),
metallicity correlations, planet and core compositions are described in \S
4-9. The radial distribution of simulated planets is compared to observational
constrains in \S \ref{sec:radial}. A broad discussion of the implications of
the results is given in \S \ref{sec:discussion}.

%
%
\section{Computational approach}\label{sec:preliminaries}

\subsection{Main assumptions}\label{sec:assumptions}

Our main assumptions are very similar to those in paper I and II. We recount
them briefly below, with more detail on the calculations to follow in \S
\ref{sec:methods} below.

\begin{enumerate}

\item We study the migration of a gas fragment in an initially massive
  self-gravitating disc around a star with mass $M_* = 1\msun$ in the last
  phase of the disc existence. The fragments are born at separation $a \sim
  100$~AU where this process is allowed by physical conditions. During our
  simulations, the mass supply from the parent molecular cloud has ceased, and
  the disc mass monotonically decreases due to accretion on the star and the
  disc photo-evaporation. This approach neglects the earlier population of
  fragments born when the star was less massive for the sake of simplicity at
  this exploratory stage.

\item Only one fragment per disc is simulated. This is a serious shortcoming
  of the calculations as simulations of self-gravitating discs show formation
  of multiple fragments forming and interacting with each other
  \citep[e.g.,][]{BoleyEtal10,ChaNayakshin11a,Meru13}. We attempt to mitigate
  for this by variations in the initial parameters of the disc and in the
  fragment's migration speed (see \S \ref{sec:methods}).

\item Planetesimals are neglected in this paper. These minor bodies are a side
  show in the context of TD, something that also happens during planet
  formation but is not a pre-requisite to formation of any of the
  planets. Moreover, in TD planetesimals are formed by the proto-planets (that
  is, massive gas fragments) and not the other way around. Specifically,
  \cite{Nayakshin10b} argued (in \S 7 of that paper) that environments inside
  massive gas fragments are both better and safer for solids to grow than the
  main body of the protoplanetary discs. Higher gas densities and self-gravity
  of the gas fragments provide high grain growth rates and protection against
  fragmenting high speed collisions, turbulence and the rapid migration of the
  fragments into the star \citep[the famous "1 metre barrier",
    see][]{Weiden80}. \cite{NayakshinCha12} showed that within the fragment,
  solid bodies with small angular momentum accrete onto the central massive
  core, whereas $\sim 1$~km or larger bodies with a larger angular momentum
  orbit the core instead. When the gas fragment is disrupted, bodies closest
  to the core remain bound to it, possibly contributing to formation of
  satellites of future super Earth or more massive planets. Bodies farther
  away from the core become gravitationally unbound from it when the fragment
  is disrupted. They however remain bound to the parent star and form debris
  rings with orbital properties not unlike the Kuiper and the asteroid belts
  \citep{NayakshinCha12}. Most importantly, there is likely much less mass in
  these fragment-produced "planetesimals" than in the Core Accretion scenarios
  of planet formation. For this reason, although planetesimals formed in
  earlier generations of disrupted fragments may accrete onto and influence
  the gas fragment evolution \citep[e.g., see][]{HelledEtal08,BoleyEtal11a},
  this process is neglected in the present paper.
  
\item Once formed by gravitational instability, fragments do NOT accrete more
  gas from the disc. \cite{NayakshinCha13} performed 3D SPH simulations of gas
  fragments embedded in massive marginally stable discs that included a
  physically motivated analytical prescription for radiative heating of the
  gas around the planet's location.  It was found that planets less massive
  than $\sim 5-10$ Jupiter masses create a hot atmosphere around themselves
  that stifles accretion of more gas onto them. More massive planets accrete
  gas rapidly and run away well into the BD regime at which point migration of
  the fragment is stalled. More 3D simulations are needed to ascertain
  (in)dependence of this result on dust opacity of the disc and other
  parameters of the model.

\item Pebbles are accreting onto the fragments at the rate (equation 5 in
  paper II) calculated by extrapolating results from \cite{LambrechtsJ12}. The
  size of the pebbles is set to 0.3 mm for simplicity, and the incoming
  pebbles are deposited into the outermost layers of the planet. 

\end{enumerate}

Readers not interested in technical details of the calculations may skip
section \ref{sec:methods} and continue directly to section \ref{sec:overview}
where results of this paper are beginning to be presented.

\subsection{Summary of numerical methods}\label{sec:methods}

We follow numerical methods explained at length in paper I and II. These are
summarised below.  We also explain several important modifications that
improve our methods.

\subsubsection{Disc evolution}\label{sec:disc_evol}

We set the inner boundary of the disc to $R_{\rm in} = 0.08$~AU, whereas the
outer boundary is set to $R_{\rm out} = 400$ AU for this paper. We use 300
radial bins, logarithmically spaced in $R^{1/2}$, which gives us the highest
numerical resolution near the inner boundary. We found that increasing the
number of radial bins to over 1000 only led to changes in the planet migration
times at the level of $\sim 10$\%. We hence chose 300 bins as acceptably
accurate for a population synthesis study: many parameters of the model (such
as viscosity parameter $\alpha$) are known with a far worse precision anyway.

The disc is initialised with a surface density profile $\Sigma(R)\propto (1/R)
(1 - \sqrt{R_{\rm in}/R}) \exp(-R/R_{\rm disc})$, where $R_{\rm disc} =
100$~AU is the disc radial length scale, with the normalisation set by the
total disc mass, $M_{\rm d}$, integrated from $R_{\rm in}$ to $R_{\rm
  out}$. The disc surface density is evolved by solving the time dependent
viscous disc equation (cf. equation 1 in paper I), but now also including the
disc photo-evaporation term to be explained below. The disc exchanges angular
momentum with the planet via a prescription switching smoothly from type I
into type II regime. The latter regime, when a gap in the disc is opened by
the planet, is modelled by the widely used "impulse approximation"
\citep[e.g.,][]{LinPap86,armibonnell02}. The type I regime (no gap in the
disc) is motivated by the results of \cite{BaruteauEtal11}, parameterised via
the planet migration timescale, $t_{I}$, given by
\begin{equation}
t_{I} = f_{\rm migr} {M_*^2 \over M_p M_{d}} {H^2\over a^2} \Omega_a^{-1}\;,
\label{time1}
\end{equation}
where $a$ is the current position of the planet, $M_d = \pi \Sigma a^2$ is
approximately the local disc mass, $\Omega_a$ is the local Keplerian angular
velocity, and $f_{\rm migr}\sim 1$ (see below) is a free parameter. The
criterion introduced by \cite{CridaEtal06} is used to switch the planet
migration between type I and type II. We emphasise that our approach conserves
angular momentum explicitly: the torque from the disc on the planet is equal
to that of the planet on the disc with the minus sign, of course.

The disc viscosity is modelled as a \cite{Shakura73} prescription, with the
$\alpha$ viscosity parameter given by the sum of a constant term, $\alpha_0 =
0.005$ and a term that depends on the local self-gravity of the disc, 
\begin{equation}
\alpha_{\rm sg} = 0.2 {Q_0^2 \over Q_0^2 + Q^2}\;.
\label{alpha_sg}
\end{equation}
where $Q$ is the local Toomre's parameter. The parameter $Q_0$, setting the
transition from a viscosity driven by gravito-turbulence
\citep[e.g.,][]{LodatoRice05} to the non self-gravitating one, is set to
$Q_0=5$ here. We found that lower values of $Q_0$ ($Q_0=2$ was used in paper
II) result in the stars with discs fraction that is too high at early times,
$t \simlt 2$ Myrs. To fit observational constraints, one would either have to
choose a much larger value of $\alpha_0$ or require implausibly high
photo-evaporation rates.

\begin{figure}
\psfig{file=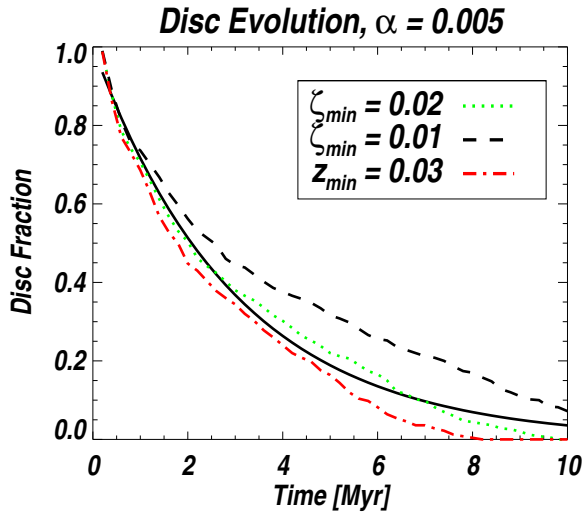,width=0.5\textwidth,angle=0}
 \caption{Fraction of stars with discs as a function of time for several
   values of the minimum disc evaporation factor $\zeta_{\rm min}$, as
   labelled in the legend. The solid line gives the curve $\exp (-t/t_{ob})$,
   where $t_{ob} = 3$ Myr. $\zeta_{\rm min} = 0.02$ provides an acceptable fit
   to the solid curve.}
   \label{fig:disc_life}
 \end{figure}

In paper II, disc photo-evaporation due to the host star's UV and X-ray
radiation was modelled as the sum of the fits to the respective
photo-evaporation rates $\dot\Sigma_{\rm in}(R)$ from \cite{AlAr07} and
\cite{OwenEtal12}. The UV ionising photon luminosity of the star was set to
$\Phi_{\rm ion} = 10^{42}$ photons s$^{-1}$, while the X-ray flux of the star
was $L_X = 2\times 10^{30}$ erg~s$^{-1}$. In this paper we employ the same
approach, but also include the external photo-evaporation rate, $\dot
\Sigma_{\rm ext}$, by UV flux from nearby massive stars at $0.1$ times the
rates parametrised by \cite{Clarke07}.  This procedure in general is not
guaranteed to satisfy observational constraints on disc dispersal,
however. Following \cite{MordasiniEtal09a}, we introduce a free parameter, $0
< \zeta_{\rm ev} $, so that the photo-evaporation rate is given by
\begin{equation}
\dot \Sigma_{\rm ev} = \zeta_{\rm ev} \left(\dot\Sigma_{\rm in} + \dot \Sigma_{\rm ext}\right)\;.
\label{sigma_dot}
\end{equation}
With this photo-evaporation prescription, the total photo-evaporation rate
from the disc is equal to $\dot M_{\rm ev} = 3.1\, \zeta_{\rm ev}\times
10^{-8}\msun$~yr$^{-1}$. As \cite{MordasiniEtal09a}, we generate a uniform
random distribution of $\log \zeta_{\rm ev}$ bounded from below by $\zeta_{\rm
  min}$ and $\zeta_{\rm max}$, and evolve a sample of $\sim 1000$ discs {\em
  without} planets. We then plot the fraction of stars with discs as a
function of time, and compare it with the observationally established
dependence, $\sim \exp (-t/t_{ob})$, where $t_{ob} = 3$ Myr
\citep{HaischEtal01}.  By experimenting with values of $\zeta_{\rm min}$ and
$\zeta_{\rm max}$, it was found that a reasonably good agreement with the
observations is obtained for $\zeta_{\rm min} = 0.02$ and $\zeta_{\rm max} =
10$ (cf. fig. \ref{fig:disc_life}). There is a certain degeneracy in this procedure. A higher value of disc viscosity 
($\alpha_0$) could be offset by reducing the
photo-evaporation rate, but our main results do not depend on this level of
detail.

\subsubsection{Planet evolution module}\label{sec:planet_mod}

The planet evolution module used in this paper is nearly identical to that
described in section 4 of paper I. In brief, we model the pre-collapse gas
fragment contraction using the ``follow the adiabats'' approach, which assumes
the fragment to be strongly convective except for the boundary, and includes
the radiative cooling of the fragment modified by the irradiation it receives
from its surroundings \citep[which can be very important, even unbinding the
  planet, see][]{CameronEtal82,VazanHelled12}.  The fragment is embedded in
the disc if it migrates in the type I regime, and is hence bathed in the disc
radiation with effective temperature of the disc mid plane.  When the fragment
migrates in type II regime, the irradiation from the central star and the
edges of the gap carved by the fragment may be important. One difference from
paper I is that we modify the intrinsic luminosity of the planet, $L_{\rm
  rad}$, by introducing a ``green house'' effect for the planet immersed into
the disc:
\begin{equation}
L_{\rm rad} = {L_{\rm iso} \over 1 + \tau_{\rm d}}\;,
\label{lrad}
\end{equation}
where $L_{\rm iso}$ is the luminosity of the fragment in isolation, and
$\tau_{\rm d} = \kappa \Sigma_a /2$ is the disc optical depth at the planet's
location. This correction is necessary because the irradiation luminosity
incident on the planet from the disc was calculated in paper I and II {\em
  neglecting} the planet's energy input to the disc around the planet's
location. Furthermore, the disc thermodynamic properties are azimuthally
averaged in our 1D approach, which additionally washes out the planet's
influence on the thermal balance of the disc in the planet's vicinity ($\sim
R_H$ or $H$, whichever is larger).  The planet's input into the disc
temperature around its location must however be taken into account since
massive planets (several Jupiter masses or more) can easily dominate the disc
energy production around the planet \citep[e.g.,][]{NayakshinCha13}. The
planet's luminosity calculated via equation \ref{lrad} is therefore used in
place of $L_{\rm iso}$ in equation (9) of paper I to determine the rate of
planet's radiative energy loss.

The dust physics used in this paper is identical to that described in sections
4.2, 4.3, and 4.6 of paper I. We use three grain species -- water,
CHON\footnote{CHON is a mnemonic acronym for a material dominated by carbon,
  hydrogen, oxygen, and nitrogen, excluding water. Its composition, material
  properties and vapor pressure are here taken to be similar to that of the
  grains in the coma of Comet Halley \cite{KruegerEtal91,Oberc04}. CHON is a
  widely used component in planet formation theory
  \citep[e.g.,][]{PollackEtal96,HelledEtal08}.}  and rocks in relative
abundances of 0.5, 0.25 and 0.25.

The total grain mass abundance is initially equal to the metallicity of the
system, $Z$. The grains grow by sticking collisions, can sediment down modulo
turbulence and convective grain mixing which oppose grain
sedimentation. Grains can also be fragmented if they sediment at too high
velocity, which in practice limits their size to a few cm at most. Grains
vaporise once the given species vaporisation temperature is exceeded. Core
assembly by grain sedimentation therefore depends sensitively on the
temperature and convection of the inner regions of the gas fragment.  The core
growth prescriptions are identical to those presented in section 4.4 of paper
I.

\subsubsection{Atmospheres bound to solid cores}\label{sec:new_atmo}

One significant modification to our treatment of the planet's internal
evolution is addition of an ``atmosphere'' around the growing massive
cores. As found by \cite{NayakshinEtal14a}, when the core reaches the mass of
a few to a few tens of $\mearth$, the weight of the gas surrounding the core
becomes comparable to the weight of the core itself. In
\cite{NayakshinEtal14a}, the structure of the core's atmosphere was calculated
assuming hydrostatic and thermal balance. Additionally, the outer boundary
conditions for the atmospheric structure calculations were fixed by joining the
atmosphere to a non-perturbed approximate analytical solution for a gas
fragment from \cite{Nayakshin10c}. Unpublished experiments done with a 1D
radiation hydrodynamics code \citep[described most recently
  in][]{Nayakshin14b,Nayakshin15a} showed that collapse of the atmosphere
around the core does {\em not} actually lead to the collapse of the whole
fragment, as assumed in \cite{NayakshinEtal14a}. The physical reason for this
is that during such a collapse contraction of the atmosphere generates a
significant compressional energy, which is then communicated to the rest of
the fragment by radiation and convection. The fragment then expands, which
lowers the density at the fragment--atmosphere boundary. As the result, the
core's atmosphere's outer layers also expand outwards, preventing further
collapse of the gas layers near the core. This ``feedback loop'' prevents a
hydrodynamical collapse of the whole fragment that was suggested to occur in
\cite{NayakshinEtal14a}. Physically, we expect a slow contraction of the
atmosphere instead at the rate regulated by the cooling of the whole fragment.

To take this physics into account, we calculate the mass (and structure) of
the atmosphere near the core at regular (short) time intervals. The atmosphere
region is defined as that around the core where the total specific gas energy,
$u - GM(r)/r < 0$, where $u$ is the specific internal gas energy, $r$ is the
radius from the core, and $M(r)$ is the enclosed mass within that radius,
including both gas and the core. If the fragment collapses due to H$_2$
dissociation then no modifications are done, since the atmosphere just becomes
the part of the gaseous envelope of the planet. However, if the fragment is
tidally disrupted, then the mass of the disruption remnant should be more massive than
just the core (which was assumed in papers I and II). Specifically, if at
disruption the mass of the atmosphere, $M_{\rm atm}$, is lower that the
critical mass for atmosphere collapse, $M_{\rm ac} = (1/3) M_{\rm core}$, then
the mass of the survived planet is given by $M_p = M_{\rm core} + M_{\rm
  atm}$. However, if $M_{\rm atm} > M_{\rm ac}$, then we expect that a
fraction of the total mass of the gas envelope would have accreted onto the
core. Since our planet formation module is not yet sophisticated enough to
calculate this fraction from first principles, we simply endow the core with
gas mass randomly distributed in $\log$ between $M_{\rm ac}$ and the total
mass of the fragment minus that of the core.  This procedure is equivalent to
a partial tidal disruption of a fragment, which typically results in planets
with masses between a few tens of $\mearth$ to $\sim 1$ Jupiter mass. We find
that partial tidal disruptions of fragments more massive than $\sim 2 \mj$ are
rare.

\subsubsection{Monte Carlo initial conditions}\label{sec:MonteCarlo}

Table 1 summarises the randomly drawn variables and their ranges used in this
paper. The approach is similar to that from paper II: for each random
variable, the distribution is uniform in the $\log$ of the parameter, and is
distributed between the minimum and the maximum values given in Table 1. For
example, the initial mass of the gas fragments is $M_0$ in Jupiter masses,
with the $\log M_0$ distributed randomly from $\log (1/3)$ to $\log 16$. Note
that the logarithmic distribution for $M_0$ translates into the $dN/dM_0
\propto 1/M_0$ dependence for the mass function of the fragments, which
corresponds well to the one observed for gas giants \citep[][finds
  $dN/dM_0\propto M_0^{-1.05}$]{US07}. This range is far broader than the one
used in paper II, where we only sampled the planets with initial masses in the
range $0.5 \mj < M_0 < 2 \mj$. We hope to make tentative connections of TD
hypothesis to the brown dwarf regime.

The disc mass is sampled between $0.075 \msun$ and $ 0.15 \msun$ of that. As
explained in paper II, the mass of the disc at which the disc is
gravitationally unstable (Toomre's parameter $Q=1.5$) is about $0.15\msun$
(cf. fig. 1 in paper II) for the chosen initial surface density profile. Lower
disc masses are considered here because 3D simulations show it is possible for
fragments to be born in less massive non self-gravitating discs due to
fragmentation induced by perturbations from fragments born at previous more
massive disc epoch \citep[e.g.,][]{Meru13}. The fragments initial locations
are sampled between 70 AU and 1.5 times that, which is reasonable since the
minimum in the Toomre parameter is reached at $a \sim 90$ for our disc.

The opacity reduction factor is set to 1 in this paper \citep[that is, the
  interstellar grain opacity of][not modified by grain growth, is
  used]{ZhuEtal09}. The planet migration factor, $f_{\rm migr}$, is sampled
between $0.5$ and 2, which is much smaller a range than the 1 to 10 range used
in paper II. As in previous papers, we do not actually model the internal
structure of the core, assuming it has a fixed density of 5 g cm$^{-3}$, and
we hence cannot calculate the core's luminosity self-consistently. To model
the core's luminosity, the Kelvin-Helmholtz contraction time of the solid
core, $t_{\rm kh}$, is defined \citep[see][]{NayakshinEtal14a}, and the core's
luminosity is prescribed as in section 4.4 of paper I. The parameter $t_{\rm
  kh}$ is varied between $10^5$ and $10^7$ years for the simulations in this
paper.

\begin{table*}
\caption{The range of the Monte Carlo parameters of the population synthesis
  calculation. The first row gives parameter names, the next two their minimum
  and maximum values. The columns are: Planets initial mass, $M_0$, in Jupiter
  masses; $\zeta_{\rm ev}$, the evaporation rate factor
  (cf. eq. \ref{sigma_dot}); $f_{\rm p}$, the pebble mass fraction determining
  the fraction of the disc grain mass in the pebbles; $a_0$ [AU], the initial
  position of the fragment; $M_{\rm d}$, the initial mass of the disc, in
  units of $\msun$; $f_{\rm migr}$, the type I planet migration factor;
  $t_{\rm kh}$, Million years, determines the luminosity of the core;
  $\alpha_d$, turbulence parameter within the fragment; $v_{\rm br}$, the
  grain breaking velocity, in m/s}
\begin{tabular}{lccccccccc}
Parameter & $M_0$ & $\zeta_{\rm ev}$ &   $f_{\rm p}$ & $a_0$  & $M_{\rm d}$ &
$f_{\rm migr}$ & $t_{\rm kh}$ & $\alpha_{d}$ & $v_{\rm br}$ \\ \hline
Min &  $1/3$ & 0.02 &  0.04 &  70  &  0.075 & 0.5  &  $10^5$ & $10^{-4}$ & 5 \\
Max &  16& 10.0  &  0.08  &  105 &  0.15  & 2 &  $10^7$   & 0.01 & 15 \\ \hline
\end{tabular}
\label{tab:1}
\end{table*}

The only exception from the uniform random parameter distribution procedure is
the host star/disc metallicity distribution, which can be constrained
observationally. It appears to be close to a Gaussian distribution in the
CORALIE planet sample \citep{UdryEtal00}. As shown by \cite{MordasiniEtal09a},
a good fit to the observations is provided by the Gaussian distribution
\begin{equation}
{dp(Z_L) \over dZ_L} = {1\over \sigma (2\pi)^{1/2}} \exp\left[ - {Z_L^2\over 2
    \sigma^2}\right]
\label{zdist}
\end{equation}
where $Z_L = $[Fe/H] = [M/H], the usual logarithmically defined metallicity,
$\sigma = 0.22$, and $dp/dZ_L$ is the probability density.  The subscript ``L''
on $Z_L$ stands for ``logarithmic'', since $Z_L \equiv \log_{10}
(Z/Z_\odot)$, where $Z_\odot = 0.015$ is the Solar metallicity
\citep{Lodders03}, e.g., the fraction of mass in astrophysical metals compared
to the total mass.

The metallicity distribution of the host stars used here is hence
observationally motivated and differs from the fixed metallicity bins used in
our previous papers. The metallicity of the disc and the initial metallicity
of the planet, $Z_0$ are equal to that of the host star,
\begin{equation}
Z_0 = Z_\odot \; 10^{Z_L}\;.
\end{equation}

 %
 %

\section{Results overview}\label{sec:overview}

We performed 20,000 fragment-disc evolution experiments for this paper. In
this section we provide an overview of the results, with further detailed
analysis to follow.

\subsection{The mass-separation plane}\label{sec:mass_sep}

\begin{figure*}
\psfig{file=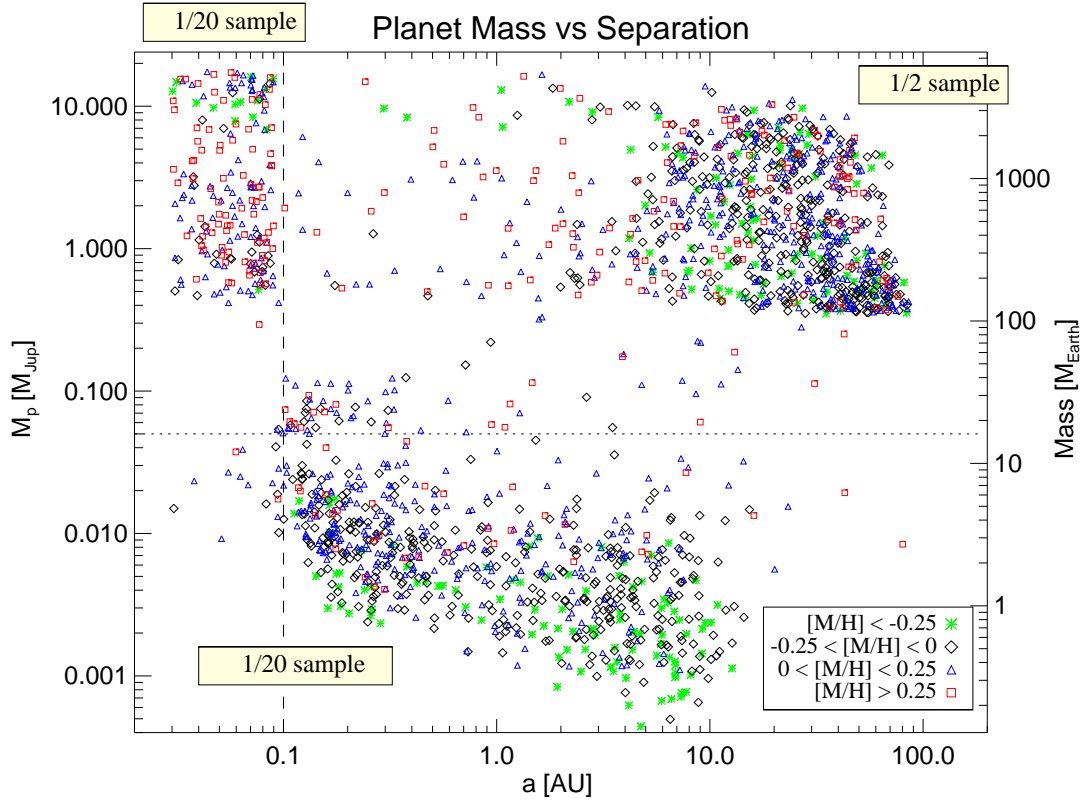,width=0.9\textwidth,angle=0}
 \caption{Simulated planets in the planet mass versus planet-host separation
   plane. The symbols and colours reflect the metallicity of the host stars,
   as detailed in the legend.  For the sake of clarity, only $1/20$th fraction
   of planets is shown to the left of the vertical line and also below the
   horizontal line. $1/2$ of all planets is shown in the right top quartile of
   the plane.}
\label{fig:scatter}
 \end{figure*}

Figure \ref{fig:scatter} presents the population of planets produced by the
end of our runs in the planet mass -- separation plane. The mass of the
planets is shown in the units of Jupiter masses on the left vertical axis, and
in the Earth mass units on the right. To improve clarity of the figure, we
reduce the number of planets (individual symbols) shown in the figure by
randomly selecting a sub-sample of the planets. The regions to the left of the
vertical dashed line (set at 0.1 AU) and below the horizontal dotted line (set
at 0.05 $\mj$) show only $1/20$th of the sample since these regions are very
strongly crowded with planets.  The remaining region -- above the horizontal
line and to the right of the vertical line -- shows $1/2$ of the sample.

The planets to the left of the vertical line are those that migrated through
our inner boundary condition (operationally, they reached very close to it,
$a=0.09$ AU). Our choice of the inner disc radius, $R_{\rm in} = 0.08$~AU, is
somewhat arbitrary. It is likely that most of these planets will be driven so
close to the star that they are either consumed by the star or are unbound by
over-heating due to the intense radiation field of the star. However, we
expect that magnetospheric interactions will cutoff the disc at different
distances from the star in different systems. This cutoff distance will also
vary with time in a given system as the accretion rate and the magnetic field
intensity vary. Due to these effects, we expect some of the planets that fell
trough our inner boundary to stop somewhere in the region between the star and
$R_{\rm in}$. For this reason we shall include a fraction of these planets in
our further analysis and hence we show them in fig. \ref{fig:scatter}. For
these planets, the separation is a uniform random variable distributed between
0.03 AU to 0.09 AU.

The colours of the symbols are used to indicate the metallicity of the parent
star as noted in the legend. For example, green symbols show the most
metal poor stars, [M/H]~$ < - 0.25$, whereas red show the most metal rich ones,
[M/H]~$ > 0.25$. One can notice the prevalence of red over green for gas giant
planets in the inner few AU, signifying the positive metallicity correlation
reported in \cite{Nayakshin15b} and paper II. The outer tens of AU gas giants
and the sub-Neptune planets below the horizontal line show much less, if any,
metallicity preference, hinting on the absence of metallicity correlation
there. These results were also found in paper II.

\subsection{Statistics of fragment destruction and survival}
 
 \begin{figure*}
\centerline{\psfig{file=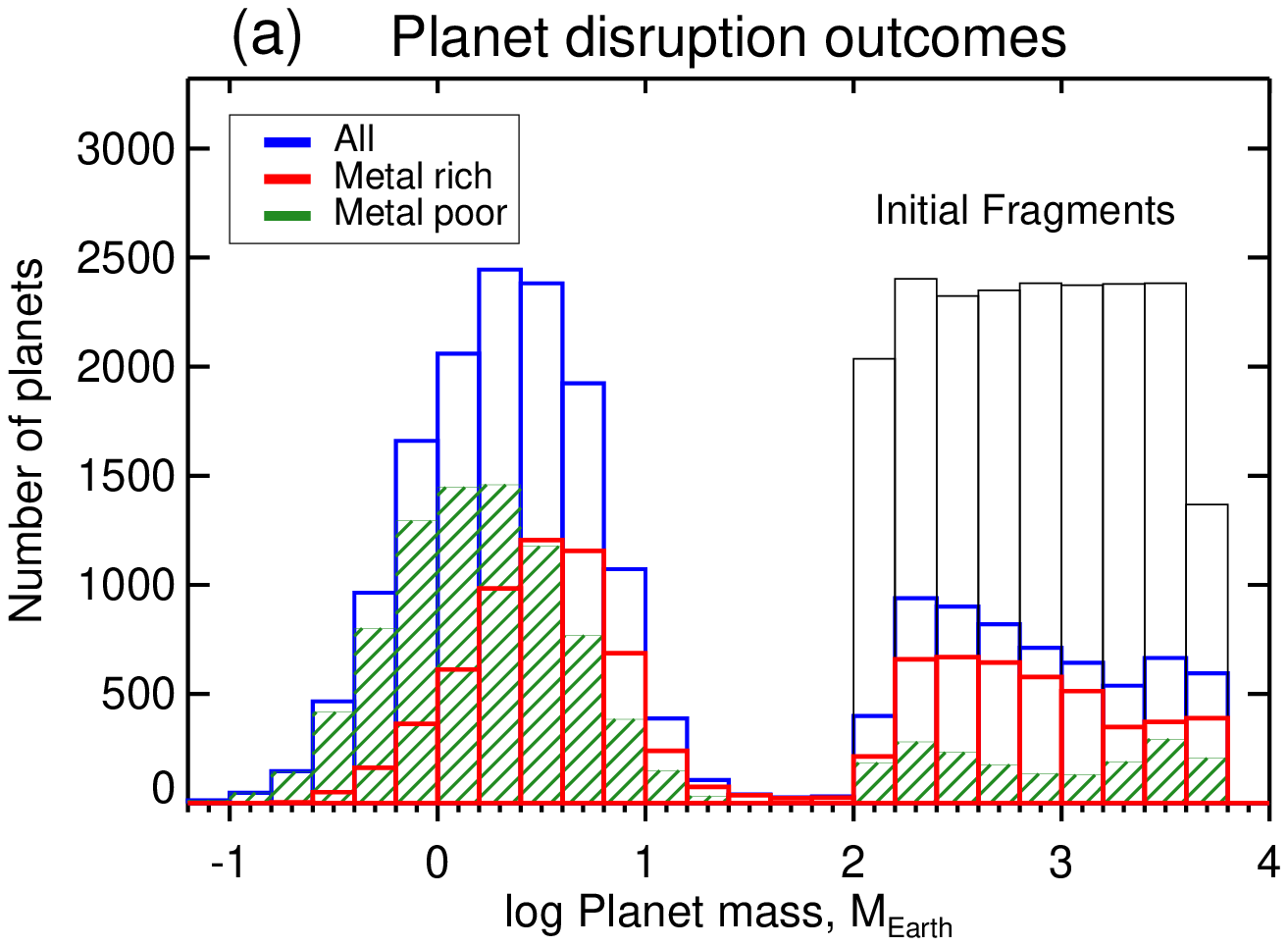,width=0.5\textwidth,angle=0}
\psfig{file=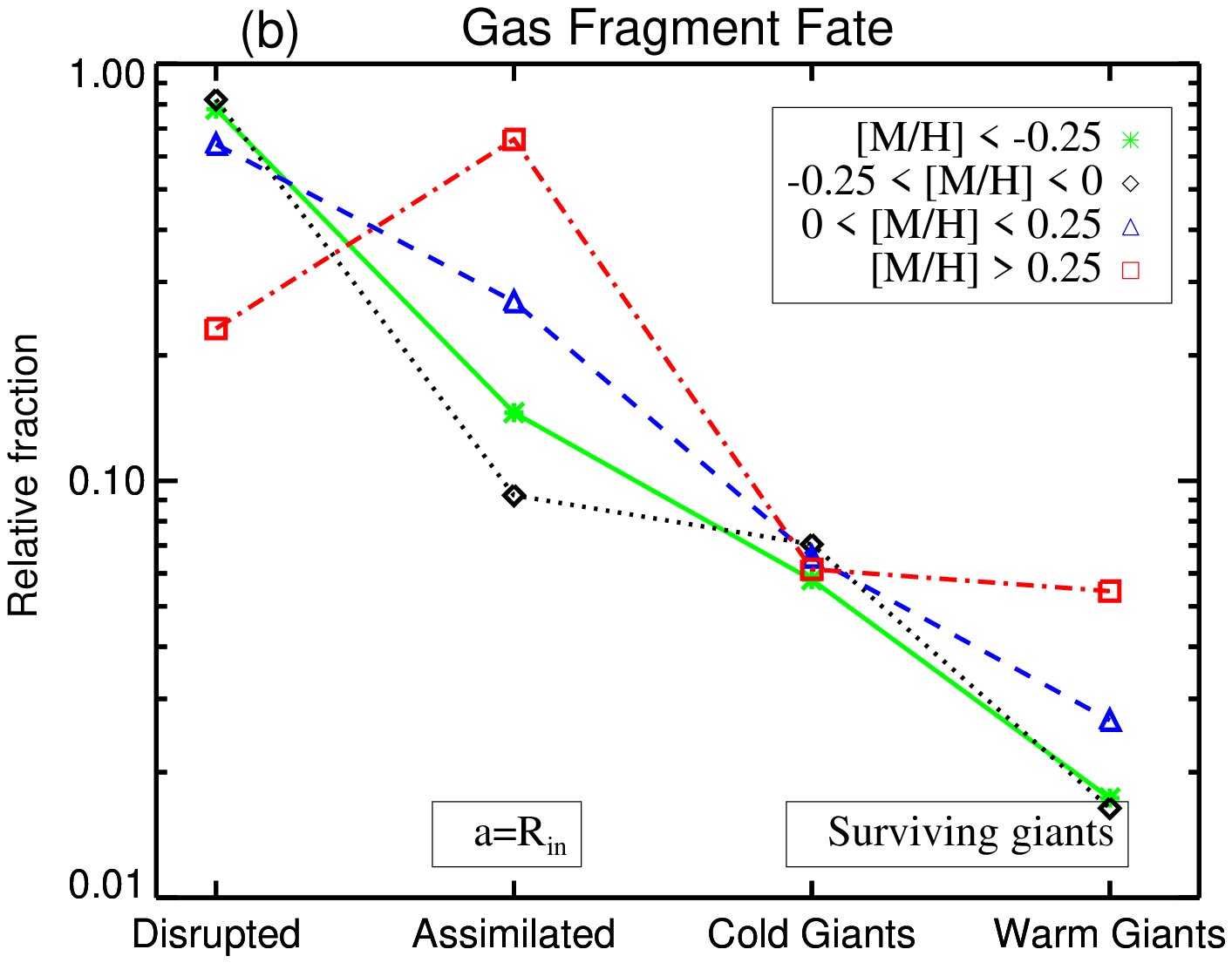,width=0.45\textwidth,angle=0}}
 \caption{(a) Comparison of the initial mass spectrum of fragments (black
   diagram) to the final distribution of planetary masses. The colours group
   the results by metallicity, with low metallicities defined as [M/H]~$ < 0$
   and high defined as [M/H]~$ > 0$. (b) The outcome of gas fragment migration
   experiments in terms of the fraction of the initial fragments that are
   tidally disrupted, pushed all the way into the star ("assimilated"),
   survive as gas giant planets at all separations, or survive as "warm
   giants" here defined as planets at separations less than $10$~AU.}
\label{fig:outcomes}
 \end{figure*}

 We shall now look at the results in a different way. All 20,000 of our
 fragments are initially born at $70 < a < 105$ AU, yet there are just some
 hundreds of planets left in the region $a > 10 $~AU by the end of the
 simulation. It is therefore clear from fig. \ref{fig:scatter} that most of
 the fragments migrate very close to the star, and are either disrupted and
 end up as low mass planets below the horizontal line, or migrate through the
 inner boundary of the disc and appear to the left of the vertical line.

 For a more quantitative analysis of this, Figure \ref{fig:outcomes}
 summarises the fate of our gas fragment migration experiments in two
 different ways. Fig. \ref{fig:outcomes}a shows the initial fragment and the
 final planet mass spectrum. The former is shown with the black histogram, and
 is uniform in mass from $M_{\rm min0} = 1/3 \mj$ to $M_{\rm max0} = 16 \mj$.
 The quantity shown in the histograms is $dN_p/dM_p \Delta M_p$, the number
 of planets found in the mass bin between $M_p$ and $M_p + \Delta M_p$. Some
 deviations from a flat line in the figure are due to random Monte Carlo
 fluctuations in the fragment mass initialisation procedure, and also $M_{\rm
   min0}$ and $M_{\rm max0}$ falling inside the histogram bins rather than
 precisely on the bin edges.
 
 The coloured histograms show the final masses of the planets after TD
 "processing" of the fragments in the disc. These are shown for all planets
 independently of where they end up in terms of their final planet-host
 separation. The populations are further broken on the metal rich bin, [M/H]~$
 > 0$ (red colour), the metal poor bin, [M/H]~$ < 0$ (green) and the total
 (blue).

We see that the distribution of final planet masses broadly divides into two
groups: the disrupted one (fragments less massive than $1/3 \mj$ or $\sim
100\mearth$) and those that avoided the disruption. The former population is
more populous than the latter, as we shall see. The most populous group of
disrupted planets are those that have masses from sub-Earth mass to $\sim 10
\mearth$. We notice that sub-Earth planets are more abundant in metal poor
discs whereas super-Earths are more abundant at higher metallicities. This
distribution however contains planets at all separations from the star and
needs to be cut into more specific radial bins to learn more, which will be
done  below.

To digest the overall results of our simulations further, we find it useful to
group the end outcome of the runs into:
\begin{enumerate} 
\item Disrupted planets -- those that were tidally disrupted anywhere
  in the disc between $R_{\rm in}$ and their birth position.

\item ``Assimilated'' planets: those that migrated all the way to the inner
  boundary of our disc, $R_{\rm in}$. Our calculations are oversimplified at
  the inner boundary, as pointed out above. In all likelihood, most of these
  fragments would be pushed closer to and possibly even into the star and thus
  be accreted (assimilated) by the star, although a minority may survive when
  the disc dissipates away.
 
\item ``Cold giants'' -- those fragments that survived tidal disruption and
  matured into a giant planet at large ($a > 10$~AU) separation from the star.
 
\item ``Warm giants'' -- those fragments that survived tidal disruption,  and,
  despite the rapid inward migration, were able to stop outside the inner disc
  boundary but within $a  < 10$~AU.
 
 \end{enumerate}
 
The frequency of these four outcomes are displayed in fig. \ref{fig:outcomes}b
separately for the four metallicity bins. Note that tidal destruction of the
fragments is the most common outcome ($\sim 80$\% of the time) for Solar
metallicity or less metal rich stars, but becomes progressively less common
for the two metal rich bins, dropping to just over $ 20$\% for [M/H]~$ >
0.25$. This result is related to the positive metallicity correlation for gas
giant planets found by \cite{Nayakshin15b} and in paper II: the higher the
metallicity, the fewer giants are disrupted, and hence more giants survive per
star.

Most of the fragments in the most metal-rich discs (red curve in
fig. \ref{fig:outcomes}b) avoid tidal disruption. However, most of them end up
migrating through the whole of the disc to $R_{\rm in}$. As already suggested,
a fraction of them is likely to survive inside $R_{\rm in}$ as "hot
Jupiter". Focusing now on the population of gas fragments that were not
tidally disrupted and not migrated through all of the disc (the last two
entries on the horizontal axis in the figure), we observe that most of these
reside at large separations from the star, $a > 10 $~AU.  The gas giant
planets in the inner few AU is hence the rarest gas giant population in our
synthetic models.  Finally, it is possible to notice a strong positive
correlation with metallicity in the fraction of giants survived in the inner
disc ("warm giants") and no such correlation in the outer disc.

%
%
\section{Hot planet mass function}\label{sec:pmf}

The planet mass function (PMF) is one of the most interesting observables of a
planet population. Figure \ref{fig:outcomes}a presented the mass function of
planets formed in our simulations at all separations. 
The observed PMF are available only for close-in planets at this point in time
as directly imaged planets are rare \citep[e.g.,][]{BowlerEtal15} and can only
sample planets more massive than $\sim 1 \mj$. We shall therefore focus our
attention on the inner part of the mass-separation diagram.

Since our model is still in the development stage, we shall not attempt a
detailed comparison between the theory and the data by performing a synthetic
``observation'' \citep[see][]{MordasiniEtal09b} of our planets. Instead, we
shall make a rather crude cut of the parameter space by carving out a ``hot''
planet sample which we define as the planets found within the inner 5~AU of
the host star plus a 10\% fraction of the "assimilated" planet group, that is,
those planets that ended up at $a=R_{\rm in}$. The latter group of planets is
included in our analysis for the reasons explained in \S \ref{sec:mass_sep}
which is also similar to the approach of
\cite{MordasiniEtal09b}. Specifically, our inner boundary condition is too
simple at this stage of the model development, yet observations, especially
transit surveys, are heavily biased towards these planets. We expect that some
of the simulated planets at $a=R_{\rm in}$ will survive in the region between
the star and $a=R_{\rm in}$.  Our 10\% choice of planets at $a=R_{\rm in}$ is
ad hoc but this choice does not influence any of the conclusions of this paper
significantly.  We also consider below a planet sample that does not
  include the assimilated planets at all.

\subsection{Massive planets}\label{sec:pmf_giants}

\begin{figure}
\psfig{file=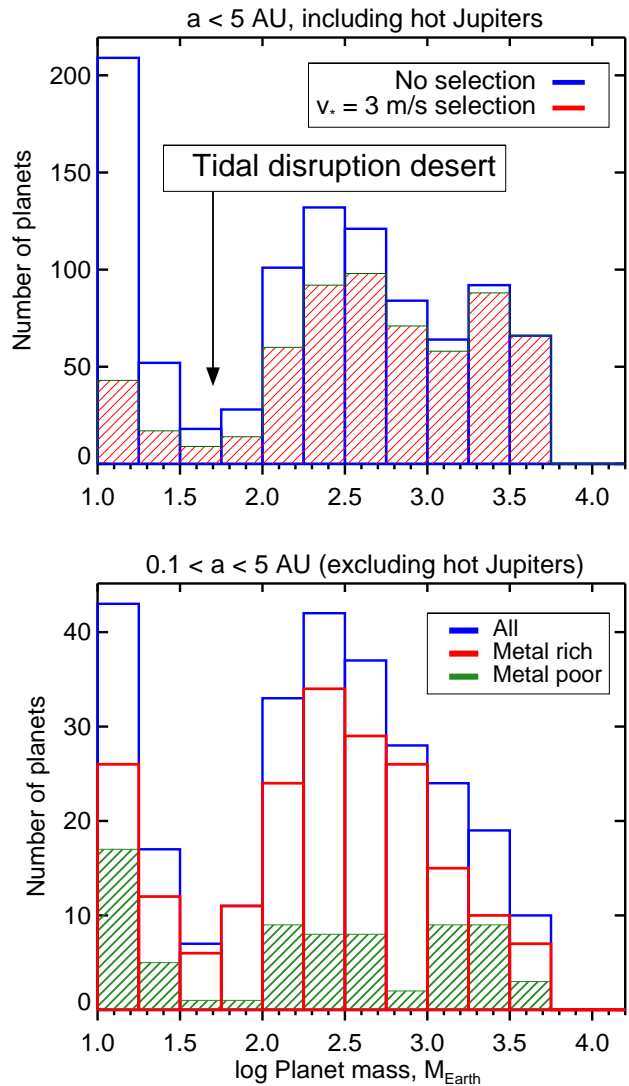,width=0.5\textwidth,angle=0}
 \caption{Planet mass function (PMF) from simulations. The top panel is the
   ``hot planet sample'' defined in \S 4, with 10\% of ``hot Jupiter'' planets
   included. The blue shows all planets whereas the red histogram shows
   planets affected by a velocity cut. The bottom panel shows planets in the
   same spatial disc region, with velocity cut imposed, but now excluding
   planets at $a < 0.1$~AU. The colours on the bottom panel separate metal
   rich and metal poor host populations. See \S \ref{sec:pmf_giants} for
   detail.}
 \label{fig:pmf_mass}
 \end{figure}

The top panel of fig. \ref{fig:pmf_mass} shows the PMF of the hot planet
sample for planets with mass greater than $10 \mearth$. We show two versions
of PMF. The histogram shown with the blue line shows all the planets in the
sample. The shaded red color histogram shows the same sample but with a
rudimentary RV-selection criterium of the stellar Dopler shift velocity
exceeding $v_{\rm det} = 3$~m/sec. To enable this selection, we generate a random
uniform distribution of $\cos i$, where $i$ is the inclination angle of the
planet's orbit ($i=0$ corresponds to viewing the star-planet system along the
rotation axis). The stellar Dopler's velocity as seen by the observer is then
\begin{equation}
v_* \approx v_K {M_p \over M_*} \sin i\;,
\label{v_*}
\end{equation}
where $v_K = \sqrt{GM_*/a}$ is the planet's circular Keplerian velocity at its
position, $a$, and $M_* = 1\msun$. We then prescribe the probability of
  detecting a planet with a given $v_*$ value as
\begin{equation} 
P_{\rm det}\left(v_*\right) = {v_*^2 \over v_*^2 + v_{\rm det}^2}\;.
\label{pdet}
\end{equation}
This is ad-hoc, but is asymptotically correct, so that all the planets
  with large $v_* \gg v_{\rm det}$ are detected, whereas none of the planets
  with $v_* \ll v_{\rm det}$ are detected. The smooth transition is more
  realistic than applying a sharp cutoff by requiring detection only for
  planets $v_* > v_{\rm det}$. The mass shown in the histogram selected by
$v_* > v_{\rm det}$ m/s is $M_p \sin i$ rather than $M_p$, of course.

The bottom panel of fig. \ref{fig:pmf_mass} shows the PMF of the simulated
planets, with the velocity cut applied, and in which no planets inside the
inner disc edge were included, so that $a \ge 0.1$~AU for all of these
planets. We also separated the population on the metal rich (red) and metal
poor (green) samples.

\subsubsection{Tidal disruption desert}\label{sec:td_valley}

One of the notable features of the simulated PMF, in both projections in
fig. \ref{fig:pmf_mass}, is the deep depression in the number of planets with
masses between $\sim 20\mearth $ and $\sim 100\mearth$. An appropriate name
for the feature is "Tidal Disruption desert'' as it is due to the
following. To the right of it, there are massive $M_p\simgt$ a few hundred
$\mearth$ gas-dominated planets that did {\em not} experience a tidal
disruption. To the left, on the other hand, are much smaller rocky core
dominated planets that are the most abundant remnants of the disruptions. The
planets inside the desert, on the other hand, are the planets that went
through a partial disruption. These planets consist of a massive solid
core dressed in a very dense gas envelope bound to the core as explained in
section \ref{sec:new_atmo}. The gas masses of these envelopes are comparable
to or a few times larger than their core masses. We find that such planets are
rare for the simulations presented in this paper.

\begin{figure}
\psfig{file=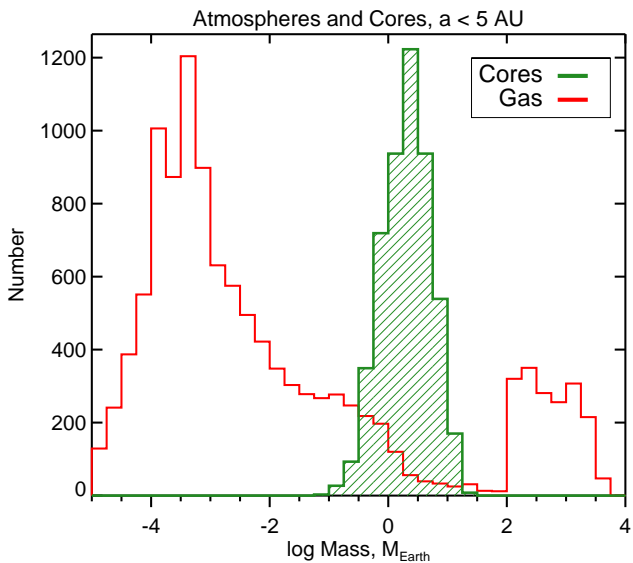,width=0.5\textwidth,angle=0}
 \caption{Mass function of bound gas in the simulations (red histogram),
   compared with that of the cores (shaded green). The latter is scaled down
   by a factor of 3 for clarity. The bound gas is either self-bound (the high
   mass peak to the right) or is attached to numerous low mass cores (low mass
   part of the red histogram), with little in between.}
 \label{fig:pmf_atmo}
 \end{figure}

To expand on this point, fig. \ref{fig:pmf_atmo} compares the mass function of
all gaseous envelopes at $R_{\rm in} < a < 5$~AU, shown in red colour, with
that of the cores (shaded green histogram). We see that gas envelopes (which
may be quite metal rich, see \S \ref{sec:atmo}) display a certain bimodality:
they are either very low mass or very high mass. The low mass atmospheres are
the majority; these are envelopes bound to low mass cores ($M_{\rm core}\simlt $
a few $\mearth$). Higher mass cores attract more massive atmospheres, but then
the number of massive ($M_{\rm core} \simgt 10 \mearth$) cores is small, and
hence the number of more massive gas envelopes plunges as well. At gas mass of
about $100 \mearth \approx M_{\rm min}$ and above there is the domain of the
undisrupted gas giants.

 While we are very confident that the Tidal Disruption desert is a robust
 feature of our model, its depth is probably not modelled reliably enough
 yet. The bound atmosphere structure is dependent \citep{NayakshinEtal14a} on
 the poorly known opacity and the equation of state \citep{StamenovicEtal12}
 of massive cores that are much hotter than present day cores of gas giant
 planets are believed to be. Additionally, our equation of state for the gas in the region
 very close to the core \citep[see][for how this may affect the critical core
   mass]{HoriIkoma11} is over-simplified. Formation of more massive
 atmospheres than found here would fill the desert somewhat, although
 evaporation of atmospheres very close to the star \citep[e.g.,][]{OwenWu13},
 not modelled here, could empty it further.

Note that Core Accretion model also predicts a ``planet desert'' at a similar
planet mass \citep{IdaLin04b,MordasiniEtal09b}, for a seemingly different but
actually nearly identical physical reason. In CA, $M_{\rm core}\sim 10\mearth$ is
the critical core mass at which point a runaway accretion of gas onto the
planet commences. Planets growing in the disc ``travel'' through the desert
quickly to reach the higher mass ($M_p \sim 1 \mj$) peak, so that the chance
of a planet stranded inside the desert is relatively low.

The lesson from this is that the structure of the planets may not always (or
perhaps even rarely?) be used to constrain the formation route of a planet
\citep[see also][]{HelledEtal13a}. In the case of the desert in the PMF,
atmospheres with masses $\sim M_{\rm core}$ are unlikely, whether the planet
is built bottom-up as in CA or top-down as in the TD. This is more of a
fundamental property of matter than a
consequence of a particular formation mechanism. It therefore appears
that some other observational diagnostic must be used to distinguish between
CA and TD models. Metallicity correlations, dependences of the PMF on the mass
of the host star, or timing of planet formation may hopefully break the
degeneracy in the future.

The observed planet mass function also shows a strong depression above
  $\sim 20 \mearth$ \cite{MayorEtal11,HowardEtal12} but it does not have the
  bump at $\sim 1$ Jupiter masses that both our and CA models predict. We
  shall study this issue further in the future, but for now we note that a
  more efficient destruction of gas giant planets than our current
  calculations would reduce their number in fig. \ref{fig:pmf_mass}, and also
  increase the number of planets in the "desert region", making the models
  more compatible with observations.

\subsubsection{Over-abundance of very massive planets}\label{sec:over}

Another feature of the theoretical PMF shown in the top panel of
fig. \ref{fig:pmf_mass} is that the number of very massive planets, $M_p\simgt
10\mj \approx 3200 \mearth$, is too large compared with the observed PMF. This
region in the PMF is however sensitive to the initial fragment mass function
shape, which we set to $dN_f/dM_f \propto 1/M_f$
(cf. fig. \ref{fig:outcomes}a). This would yield a flat PMF histogram.  The
distribution of initial fragment masses is not well constrained theoretically,
and it may well be steeper than the one we used here.

Furthermore, we also made the simplifying assumption (iv) in \S
\ref{sec:assumptions} that fragments do not accrete gas. It is possible that
the most massive fragments do accrete gas rapidly \citep[cf.  simulations
  of][]{NayakshinCha13,StamatellosH15}, so that they ``run away'' into the
regime of low mass stellar companions \citep{SW08}, which would remove them
from our theoretical PMF. 

{\bf To expand on this, we note that the maximum available mass budget of gas
  to be accreted on a fragment is the mass of the disc, which is $0.15\msun$
  (see Table 1) for the simulations presented here. However, this is a
  limitation of our set-up rather than a physical one. We here investigate
  only one fragment per disc, which makes most sense when the disc
  self-gravity is in its dying phase, and hence the disc mass is
  moderate. Real systems may however form gas fragments earlier on, in class 0
  stage, when the protostar is only a fraction of its central mass, and when
  the disc is still being fed from the protostar's natal molecular
  envelope. If there is a ``runaway'' accretion for more massive gas fragments
  as discussed in \cite{NayakshinCha13}, then these fragments may grow to
  massive BDs and even low mass stellar companions. There is hence a potential
  connection of TD theory to low mass stellar companions, which may be another
  way to test it. This is however just a speculation in this point; gas
  accretion onto the fragments must be included in the calculations properly
  for reliable theoretical predictions.

}

\subsection{Core-dominated planets}\label{sec:pmf_cores}

We now focus on core-dominated planets. We select the planets less massive
than $M_p < 10^{1.5}\mearth$, slice the resulting
sub-sample into three spatial bins, and plot their PMF in figure
\ref{fig:pmf_cores}. The histogram in blue shows the population of cores
closest to the host star, in $a < 0.5$~AU region.  The red histogram covers
the region $0.5 < a < 5$~AU, and the green histogram shows cores located
further out.

\subsubsection{No pool of close-in very low mass cores}\label{sec:no_pool}

Fig. \ref{fig:pmf_cores} shows that the inner 0.5~AU region contains very few
cores less massive than $1 \mearth$.  The absence of sub-Earths is due to the
fact that tidal disruptions producing such low mass cores typically occur
further out, at a few to 10 AU region. The low mass cores migrate slowly since
the migration rate is proportional to the core's mass,
cf. eq. \ref{time1}. Most of them therefore do not make it inside the inner
0.5~AU region.

This result will probably be weakened to some degree by more realistic
calculations that would include previous generations of fragments during
earlier disc phases. Cores located in the inner few AU may get locked into
resonances \citep[e.g.,][]{Paardekooper13} with gas fragments migrating from
the outer disc and hence be pushed inward along with the fragment more rapidly
than they could migrate by themselves. If the fragment is then tidally
disrupted and the gas is consumed by the star, a second low-mass rocky core
will be left behind. If observed at the present time, the system would have no
obvious record of a past outer massive fragment's existence. Additionally,
once the gas disc is removed, the N-body interactions of cores with each other
may bring more low mass cores into the innermost region.

Despite this, the rollover of the core mass function at low masses is a
general result of our model for all separations
(cf. fig. \ref{fig:pmf_cores}), although the core mass at which the rollover
occurs decreases with increasing separation. Therefore, while the exact core
mass at the rollover is subject to further modelling, its existence is a
robust result.

We emphasise the difference from the Core Accretion way of building cores
here. In CA, {\em all} of the solid mass is initially in tiny bodies
(planetesimals). In the early stages of planetary growth, most of the mass is
in the low mass cores and ``embryos'' with mass well below $1\mearth$. The PMF
from CA calculations thus shows a strong peak at low masses, $M\simlt 1
\mearth$, see fig. 3 in \cite{MordasiniEtal09b} and fig. 2 in
\cite{MordasiniEtal12}. There is no physical reason to have such a divergence
towards smaller masses in TD theory for planet formation since the cores grow
by accretion of grains and not by collisions of numerous solid ``embryos''.

We note that this prediction of the model is consistent with {\em Kepler}
observations \citep{HowardEtal12} that find that the frequency of planet
occurrence (corrected for observational biases) does not diverge and instead
drops towards the smallest radius planets ($R_p \sim R_\oplus$). 

\subsubsection{A smooth PMF with no local minima at a few $\mearth$}\label{sec:no_minima}

Another interesting distinction between our calculations and that of CA theory
is that some of the latter predict that the cores can be composed of two
separate populations -- the inner smaller rocky cores and the outer more
massive icy cores. Some of CA models (e.g., see figs. 8 and 13 in
\cite{MordasiniEtal09a} and fig. 3 in \cite{MordasiniEtal09b}) show a
significant dip in the PMF in-between these two populations, at $M_{\rm
  core}\sim$ a few $\mearth$. The fact that the low mass cores are rocky and
the high mass ones are icy is best seen in figures (especially 5, 7 and 11) in
\cite{AlibertEtal13}.

As will be seen later, our cores, including the most massive ones, are
rock-dominated, and therefore there is no physical reason for two separate
populations to exist in the PMF of the cores. It is hence a smooth function of
$M_{\rm core}$ (fig. \ref{fig:pmf_cores}).

\subsubsection{A rollover above $\sim 10\mearth$}\label{sec:massive_cores}

\begin{figure}
\psfig{file=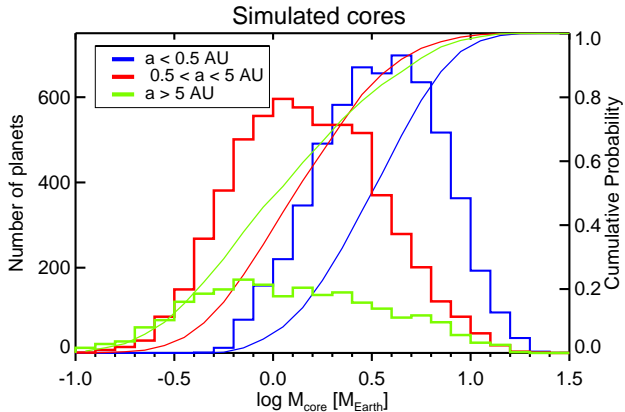,width=0.5\textwidth,angle=0}
 \caption{Planet mass function (PMF) of cores from simulations in three
   different spatial regions as explained in the legend. The vertical scale on
   the right shows the cumulative probability distribution for the same
   curves. See \S \ref{sec:pmf_cores} for detail.}
 \label{fig:pmf_cores}
 \end{figure}

The population of cores shown in fig. \ref{fig:pmf_cores} in the inner
$0.5$~AU is dominated by super Earths of mass $\sim 5\mearth$.  There are very
few cores more massive that 10$\mearth$. As noted in \S \ref{sec:planet_mod},
in TD cores grow by grain sedimentation, and this process is limited by the
rate at which the grains can sediment without fragmenting in high speed
collisions, and also by the convective mixing of the grains which tends to
bring the grains from the centre back into the outer regions of the fragment
(cf. paper I and also \cite{HB10}).  Massive cores are very luminous in our
model. This amplifies convection in the inner parts of the fragment,
forming a kind of a feedback loop that limits the rate at which the cores
grow.

Therefore, the absence of more massive cores in our models appears to be due
to an insufficient time during which the conditions within the fragments are
conducive for core growth. For a rapid core growth, the fragment must be (a)
dense enough to allow an efficient grain growth and sedimentation, (b) but not too hot
for the grains to vaporise and for the fragment to collapse into the second
core when the core will be buried under all the weight of the gas. The core
growth time window is also limited by the lifetime of the fragment from its
birth to the moment it is tidally disrupted.

\begin{figure}
\psfig{file=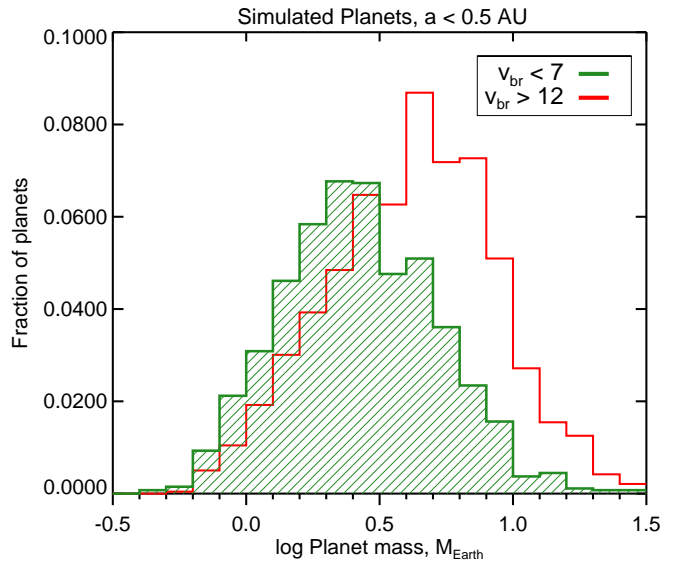,width=0.5\textwidth,angle=0}
 \caption{Mass function of low mass planets in the inner 0.5 AU for low and
   high grain breaking velocities, as specified in the legend ($v_{\rm br}$ is
   in units of m/s). Unsurprisingly, more massive cores are assembled within
   the fragments if grains can sediment without fragmenting at higher
   velocities.}
 \label{fig:pmf_vbr}
 \end{figure}

The precise value of the upper rollover mass in the core-dominated part of the
PMF does depend on parameters of the models and the atmosphere presence for
the most massive cores. To explore this issue, we slice the sample of planets for $a <
0.5$~AU into two sub-population for which the grain breaking
velocity, $v_{\rm br}$, is between $5 < v_{\rm br} < 7$~m~s$^{-1}$ and $12 <
v_{\rm br} < 15$~m~s$^{-1}$, and plot the resulting mass function in
fig. \ref{fig:pmf_vbr}. The higher breaking velocity sub-population naturally
has more massive cores because the grains can sediment at higher velocity.
This shows that the exact shape of the mass function near $M_p \sim 10-30
\mearth$ is not well constrained at the moment. Additionally, it is possible
that accretion of gas or pebbles after the tidal disruption of the fragment
would increase the mass of the planets shown in fig.  \ref{fig:pmf_cores}
somewhat.

The red histogram in fig. \ref{fig:pmf_cores} shows that low mass planets
(mainly bare cores) become more dominant at larger separations.  This is due
to the fact that high mass cores migrate inward much more than do low mass
cores. The green histogram indicates that there are relatively few cores
beyond 5 AU. This result is however sensitive to the migration prescription
and the speed with which the most massive cores are assembled; fig. 4 in paper
II shows a much larger population of cores formed at tens of AU from the host
star.

%
%
%
\section{Planet metallicity preferences}\label{sec:Zcor}

We shall now analyse how the host star metallicity influences the likelihood
of planet formation of different types. To reiterate the setup described in \S
\ref{sec:MonteCarlo}, our metallicity distribution is a Gaussian (equation
\ref{zdist}) with the mean [Fe/H] = [M/H] equal to zero and the dispersion
around that $\sigma = 0.22$, motivated by the CORALIE planet sample.  The
nature of the resulting dependences turn out to vary from a region to a region
in the planet's mass-separation plane; thus we shall study these regions
separately.

\subsection{Moderately massive giants vs super Earths in the inner 5 AU}\label{sec:SE}

 We now analyse the metallicity correlations in the hot planet sample (see \S
 \ref{sec:pmf_giants}: the planets found within the inner 5~AU of the host
 star plus a 10\% fraction of the "assimilated" planet group, that is, those
 planets that ended ended up at $a=R_{\rm in}$).  From this sample we then
 select super Earths defined as planets in the mass range $2 \mearth < M_p <
 15 \mearth$, and moderately massive giants as gas giants with mass between at
 least $100 \mearth \approx 0.3 \mj$ and $5 \mj$. More massive giants are rare
 in the observed samples. There is also a physical reason to look at the
 metallicity correlations of the more massive giants separately, as we will
 see in \S \ref{sec:BDs}.

\begin{figure}
\psfig{file=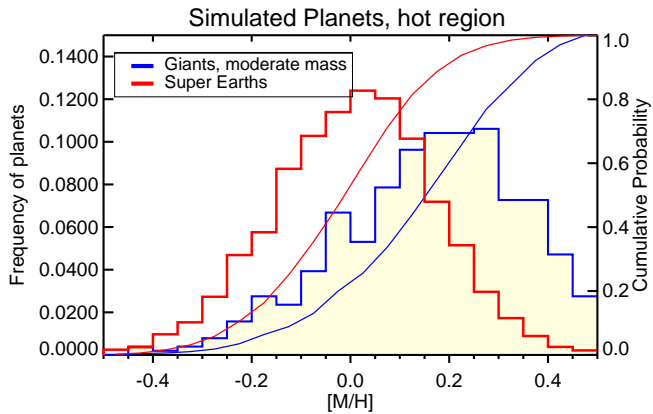,width=0.5\textwidth,angle=0}
 \caption{The fractional distribution of gas giant planets (blue filled-in
   histogram) and super Earths (red histogram) over metallicities of the host
   stars. Only planets in the inner 5 AU are selected. The curves of the
   respective colours show the cumulative distributions of same planet
   groups. Note strong positive metallicity correlation for gas giants and
   absence of such for super Earths. See text in \S \ref{sec:SE} for more
   detail.}
 \label{fig:Z_se}
 \end{figure}

Fig. \ref{fig:Z_se} shows how these two groups of planets are distributed over
the host star metallicities. The blue line histogram shows the giants, whereas
the red histogram shows the super Earths. The curves of the same colour show
the respective cumulative distributions with the scale on the right vertical
axis. If our theory were insensitive to the metallicity of the host star,
  then we would expect to see a Gaussian-like distributions centred on zero
  in fig. \ref{fig:Z_se}. However,  it is evident that gas giants appear
preferentially around metal-rich hosts, whereas super Earths are spread about
the mean metallicity roughly symmetrically.  Also note that the frequency
  of finding a gas giant {\em per star of a given metallicity} keeps
  increasing with [M/H] up to the highest value of 0.5 shown in the
  figure (see fig. \ref{fig:Znumb} below). The reason why the gas giant
  histogram goes down in fig. \ref{fig:Z_se} is that there are very few stars
  with metallicities larger than [M/H]=0.3.
 
As explained in \cite{Nayakshin15b} and papers I and II, the positive
metallicity correlation for giants is due to pebble accretion rate onto the
fragment being larger at higher metallicities, which leads to the fragment
contracting faster. Faster contraction implies fewer tidal disruptions, and
hence more gas giants surviving at higher metallicities. The physical reason
for pebble accretion being so influential in controlling the rate of the
fragment's contraction is that pebbles bring in additional mass into the
fragment (but not kinetic energy since they sediment onto the cloud gently or
else they break up in high speed collisions). Gas clumps dominated by
molecular hydrogen turn out to be very sensitive to addition of mass in this
way and collapse when the mass of the extra metals reaches $\sim 5$\% to $\sim
20$\% of the initial fragment mass. This picture also has implications for
metal overabundance in giant planets as a function of their mass as we shall
see in \S \ref{sec:z_inside}. The more massive the fragment is, the hotter it
is at birth, the less pebbles is required to bring it to the H$_2$
dissociation instability point (when the central fragment's temperature is
$\approx 2000$~K).

 The metallicity correlation for super Earths turns out to be a much subtler
 matter. Fig. \ref{fig:Z_se} shows that high metallicity environments are no
 more preferable for formation of cores in the mass range $2 \mearth < M_p <
 15 \mearth$ than low metallicity discs. In paper II, two explanations for
 this result, both working in the same direction, were proposed: (a)
 saturation in the mass of the most massive cores within the fragment at high
 metallicities, and (b) disruption of high metallicity gas
 fragments being too rare. The latter effect is important because super Earths
 are the remnants of these disruptions: no disruptions, no super-Earths.

\begin{figure}
\psfig{file=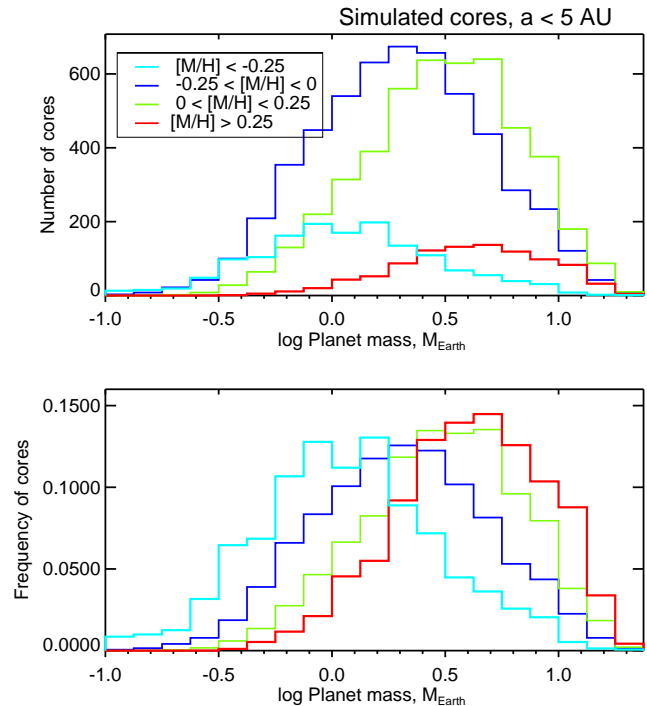,width=0.5\textwidth,angle=0}
 \caption{Number (top panel) and fractional (lower panel) mass distributions
   for cores found in the inner 5 AU for four different metallicity
   groups.}
 \label{fig:PMF_seZ}
 \end{figure}

We now investigate the relative importance of these two effects. The top panel
of Fig. \ref{fig:PMF_seZ} shows the mass function of cores found in our
simulation in the inner 5 AU for four metallicity bins as specified in the
legend. This panel shows that most of the super Earths form at metallicity not
too dissimilar from the Solar metallicity (the blue and the green
histograms). The curves are normalised by the total integral giving the total
number of cores, of course. The bottom panel of the figure shows the
distribution of frequency of planet masses for the curves in the top panel;
the histograms on the bottom are normalised on unity for each metallicity bin.
It is clear from the bottom panel of fig. \ref{fig:PMF_seZ} that the mean mass
of the core increases with metallicity rapidly while [M/H]~$ < 0.25$. However,
the mean core mass increases less between the two metal rich bins. There is
thus indeed a certain saturation in the cores mass growth at high
metallicities, although the mean core mass still increases with [M/H]~. Thus
effect (a) is not actually that strong.

The main driver of the poor correlation between super Earths and host's
metallicity in our models turns out to be (b), e.g., that the number of tidal
disruptions is just too low at high [M/H], yielding too few cores (although on
average they are more massive than the cores at lower metallicities). Figure
\ref{fig:Znumb} shows the frequency of the fragment becoming a given planet
type as shown in the legend versus metallicity of the host star. In
particular, the black connected diamonds show the frequency of obtaining a
super Earth in the inner disc. The red crosses show the frequency of the
fragment migrating through all of the disc to $r=R_{\rm in}$ without being
disrupted. Notice the anti-correlation between such un-disrupted fragments and
the number of super Earths. In particular, at the highest metallicities, as
much as $\sim 70-80$\% of our initial fragments manage to collapse and migrate
all the way in, avoiding tidal disruption. This of course must mean that fewer
core-dominated planets are made in this environment because there are fewer
tidal disruptions.

\begin{figure}
\psfig{file=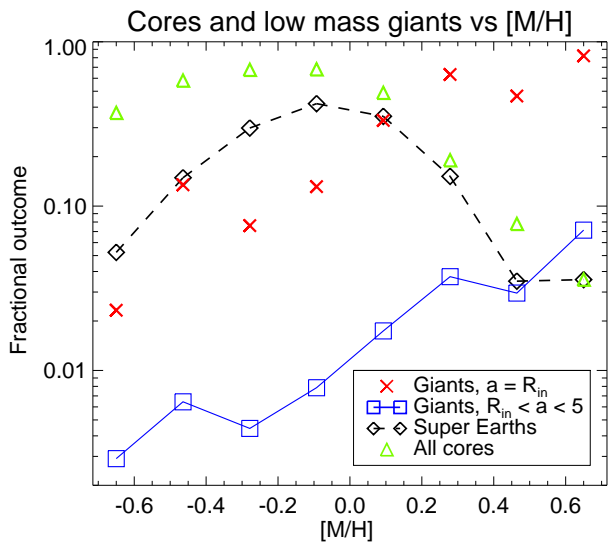,width=0.5\textwidth,angle=0}
 \caption{Frequency of a fragment becoming a planet of a given type, as shown
   in the legend, as a function of the host star metallicity. Note that at the
   highest metallicity bins, most fragments are able to migrate in to
   $a=R_{\rm in}$ and hence only a few super Earths are made. Also note
     that it is the blue square line that should be compared with Fischer \&
     Valenti (2005), since this line shows the frequency of finding a gas
     giant per star of a given metallicity, rather than that in the stellar
     population as a whole, which is presented in fig. 8.}
 \label{fig:Znumb}
 \end{figure}

Summarising these findings, we conclude that Super-Earths do not correlate
with metal abundance of host star in our models {\em because gas giant planets
  do}. The presence of a gas giant planet indicates that its predecessor, a
pre-collapse molecular fragment, was successful in avoiding the tidal
disruption. Since gas giants are most frequent at high metallicities, gas
fragment disruptions must be much less ubiquitous at high [M/H], and hence
there should be fewer Super-Earths.  On the other hand, we note that the mean
mass of the cores does increase with metallicity. Thus, Super-Earths made by
TD model may be said to correlate in mass (somewhat weakly at high core
masses) but not in numbers with the metallicity of the host star.

 The histogram distribution for our simulated planets shown in
 fig. \ref{fig:Z_se} looks quantitatively similar to fig. 2 in
 \cite{BuchhaveEtal12}, even though we did not make any attempt to arrive at
 such an agreement specifically.

\subsection{Cold giants are metal insensitive}\label{sec:cold_giants}

 \begin{figure}
\psfig{file=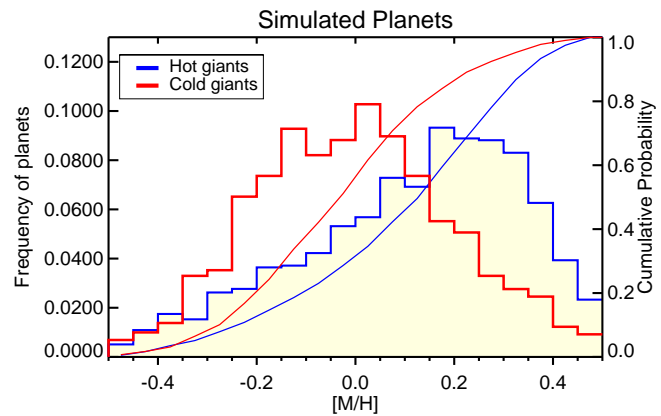,width=0.5\textwidth,angle=0}
 \caption{Same as fig. \ref{fig:Z_se} but now comparing the metallicity
   distribution of gas giant planets in the inner 5 AU (hot giants) with those
   in the "cold" region, $a> 10$~AU (red histogram). Note that cold giants
   show no metallicity preference.}
 \label{fig:Z_hotcold}
 \end{figure}

We now compare the metallicity correlation for all of the gas giant planets
$M_p > 100 \mearth$ in the hot sample of \S \ref{sec:pmf_giants} with the same
mass range planets located further out, at $R > 10$~AU, which we name ``cold
giants''.

Figure \ref{fig:Z_hotcold} compares the metallicity distributions of the hot
and cold giants. As was the case with Super-Earths discussed in \S
\ref{sec:SE}, the cold giants show little metallicity preference. The mean of
the hot giant distribution is at [M/H]~$=0.12$, clearly showing a
significant positive metallicity preference, whereas for the cold giants we
find mean[M/H]~$ = -0.01$.

The interpretation of this result is straightforward. Most of the fragments
that stayed behind 10 AU did not come close enough to the parent star to
experience tidal forces sufficiently strong to rip the fragments
apart. Therefore, these planets survive irrespectively of how much pebbles was
accreted. They are therefore insensitive to the metal content of the parent
discs.

\subsection{Most massive giants are metal insensitive}\label{sec:BDs}

Finally, we come to the most massive planets studied in this paper. A visual
inspection of the top left hand corner of Figure \ref{fig:scatter} shows that
metal poor massive giants ($M_p \sim 10 \mj$) are about as abundant as metal
rich planets of same mass. This appears true for planets at the inner disc
edge and also in the inner few AU, so it may be general for the whole hot
sample of the most massive planets.

\begin{figure}
\psfig{file=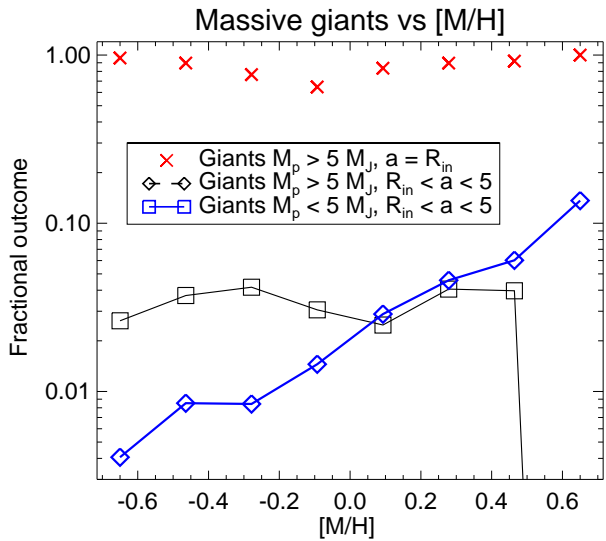,width=0.5\textwidth,angle=0}
\caption{Similar to fig. \ref{fig:Znumb}: frequency of the initial fragments
  becoming a planet in the three mass-separation ranges defined in the
  legend. Note that low to moderate mass giants (blue line) correlate
  positively with metallicity, whereas high mass giants do not.}
\label{fig:BDs}
 \end{figure}

Figure \ref{fig:BDs} shows that this is indeed the case.  Similar to
fig. \ref{fig:Znumb}, this figure shows how frequency of an initial fragment
becoming a given type of planet changes with the host star metallicity. Three
groups of gas giants is considered in fig. \ref{fig:BDs}.  The fraction of
survived moderately massive giants (the blue diamonds curve) is a rapidly
increasing function (roughly a power law $\propto Z^{5/4}$, where $Z = Z_\odot
10^{[M/H]}$), reflecting the strong positive correlation for the hot moderately
massive giants discussed in \S \ref{sec:SE}. 

The black squares curve in fig. \ref{fig:BDs} shows the survival fraction for
more massive giants, $M_p > 5 \mj$. We see that the probability of these
massive fragments surviving the disc migration phase inside the hot region is
actually flat with metallicity. Further to that, the red crosses show the
fraction of the initial fragments in the massive sample that migrated all the
way to the innermost disc radius. The curve has a depression at metallicities
just below the Solar value, which is to say that fragments around stars of
nearly Solar composition are the most likely to be tidally disrupted before
they arrive at $a=R_{\rm in}$.

For the most massive fragments, therefore, very high or very low metallicity
environments are actually preferable (although not by much, e.g., less than by
a factor of 2). This bimodality of metallicity preferences of high mass giants
forming in the context of TD was hinted on in \cite{Nayakshin15a}, where it
was shown that gas giant planets contract and collapse in one of two ways. The
most well known way, studied in literature on giant planet evolution since
\cite{Bodenheimer74,BodenheimerEtal80}, is by radiative cooling, when the
planet radiates its excess thermal energy away before collapsing. The second
way is by metal loading via pebble accretion, in which case it is the
increasing weight that leads the planet to collapse. The radiative cooling
time of the fragments is a strongly decreasing function of the fragment's mass
\citep[cf fig. 1 in][]{Nayakshin15a}. In section 7 of that paper, it was
suggested that even at large grain opacities, that is not modified by grain
growth, high mass fragments are capable of contracting faster by radiative
cooling than by pebble accretion.

The radiative cooling rate of the fragment decreases with grain opacity, which
is directly proportional in our model to the total mass of metals in the
planet minus the mass of the core. Therefore, at higher metallicities the
radiative cooling channel for collapse of massive fragments is no longer
efficient. However, pebble accretion does help at the highest metallicities,
explaining why the curve shown with the red crosses recovers to almost unity
at high [M/H].

Summarising this discussion, the metallicity correlation for the highest mass
giant planets is essentially a sum of two correlations: (i) an
anti-correlation for the radiatively cooling fragments predicted by
\cite{HB11} and (ii) the positive correlation due to pebble accretion
\citep{Nayakshin15a}. The reason why the anti-correlation becomes important
only for high mass $M_p \simgt 5 \mj$ planets and not lower mass planets is
that for lower mass planets the radiative cooling channel is not important at
all compared with the pebble accretion, even at very low metallicities.

One caveat must be emphasised here. We do not include accretion of gas onto
pre- or post-collapse giant planets (see point (iv) in \S
\ref{sec:assumptions}). It is conceivable that some of the observed massive
$M_p \simgt 5\mj$ planets could have started from gas fragments of lower mass
(even in the context of TD). If this is the case then those planets would
correlate with metallicity positively since their progenitor fragments
did. This would dilute the effects we discussed here, but we are not yet able
to quantify by how much.

\subsection{Metallicity preferences, simulation vs
  data}\label{sec:sims_vs_data}

\begin{figure}
\psfig{file=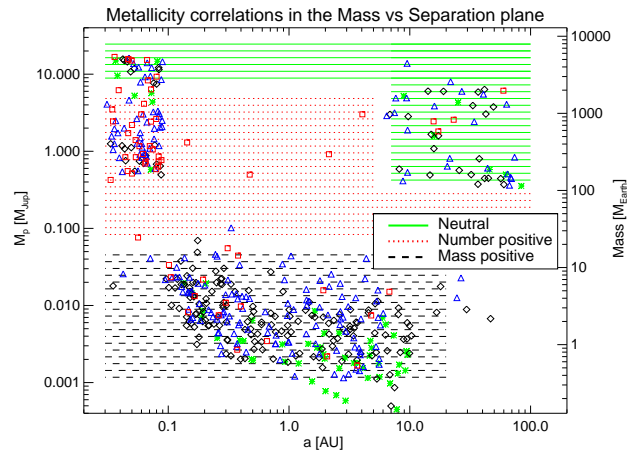,width=0.51\textwidth,angle=0}
\caption{The planet mass versus host separation plane, with 1/40th of the
  simulated planets shown, shaded by the type of metallicity correlation in
  the respective region. ``Number positive'' and ``Mass positive'' are
    correlations in which the frequency of planets or the only the mass of the
    planet correlates with the star's metallicity, respectively. See text in
  \S \ref{sec:sims_vs_data} for explanation. The symbol colour meanings are
  same as in fig. \ref{fig:scatter}.}
\label{fig:Zcorr_plane}
 \end{figure}

In the previous sections we presented theoretical predictions on how planets
correlate with the host star's metallicity in a number of regions in the planet
mass -- host star separation plane. For convenience, these are summarised
graphically in fig. \ref{fig:Zcorr_plane} in which we show a small subset of
the simulated planets ($1/40$th fraction of the total, uniformly and randomly
selected for the whole plane) overlay-ed by three shaded regions. These are:

\begin{enumerate}
\item 
In the top green shaded region, planets are insensitive to the host star's
metallicity. For cold giants (\S \ref{sec:cold_giants}) this is because these
planets do not experience strong tidal forces at their far away
locations. They hence eventually collapse independently of the presence of
pebbles in the host disc. For hot very massive planets, a second collapse
route -- by radiative cooling -- becomes available, which ``saves'' the
fragments from tidal disruptions at low metallicities (\S \ref{sec:BDs}).

\item In the red region, pebble accretion is most effective at preventing
  tidal disruptions at high metallicities, so there is a strong positive
  metallicity correlation (\S \ref{sec:SE}). Although we did not analyse this
  specifically, there is also a positive metallicity correlation for partially
  disrupted planets, e.g., with masses approximately between that of Neptune
  and Saturn, because these planets have very massive cores, and
  massive cores are assembled preferentially at high metallicity.

\item In the black shaded region, the metallicity correlation is
  complicated. If measured within a certain mass window, there is no clear
  correlation for the {\em number} of cores versus [M/H]. For example, cores
  with mass $M_{\rm core} \approx 3 \mearth$ are relatively rare at low
  metallicities (since in this case most cores are less massive than that, see
  fig. \ref{fig:PMF_seZ}), then abundant at $\sim$ Solar metallicity, and are
  rare again at high metallicities, since most cores then become more massive
  than $3 \mearth$ . However, there is an average core {\em mass correlation} with
  metallicities: the higher the metallicity, the more massive an average
  core. We called this type of correlation "number neutral but mass positive"
  in \S \ref{sec:SE}.

\end{enumerate}

\begin{figure}
\psfig{file=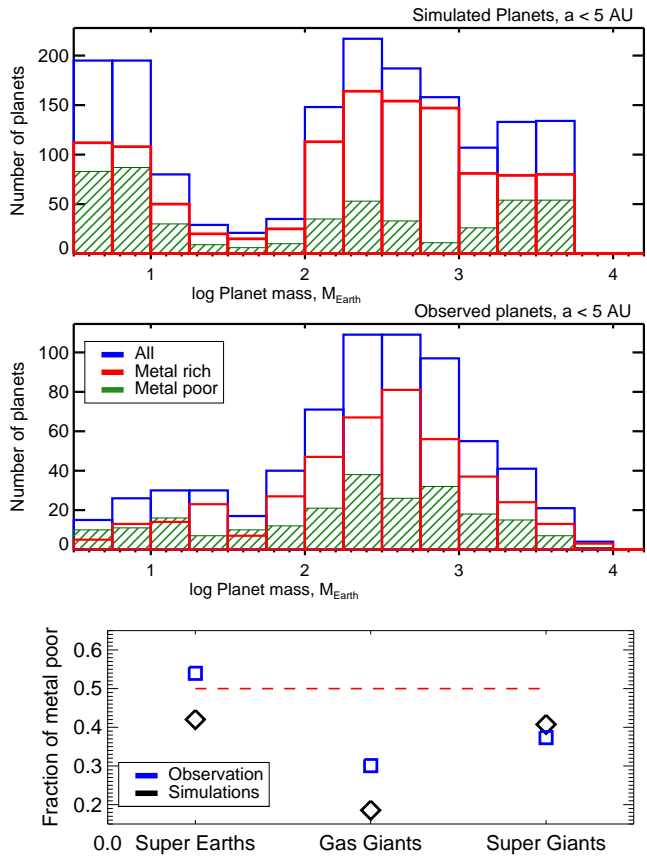,width=0.5\textwidth,angle=0}
\caption{Mass function of simulated (top) and observed (middle) planets,
  sliced into the metal rich and metal poor samples. The bottom panel shows
  the fraction of planets that are metal poor for the three selected groups of
  planets, for both observation and simulation. No attempt to include errors
  or observational biases in this figure was made.}
\label{fig:sim_obs}
 \end{figure}

It is tempting to compare these theoretical predictions to the
observations. Observations of planets in the cold region are too incomplete
for us to attempt a comparison, hence we limit our attention to the inner 5 AU
region of the simulated planets. As we wish to compare this to observations,
we included only planets with the stellar reflex velocity $v_* > 2$ m/s. This
preferentially removes the planets less massive than $\sim$ tens of $\mearth$
from the simulated sample.

 We have no access to a uniformly selected observed planet sample for the
 broad mass range to investigate metallicity correlations of the observed
 planets. Therefore, we only attempt a rather simple comparison here. We
 assume that metal-rich and metal-poor planets of same mass are affected
 roughly similarly by detection biases, and use the planets in the {\it
   exoplanets.eu} sample \citep{SchneiderEtal11}.  We select planets with
 measured masses and host-star separations less than 5 AU, and with the host
 star masses between $0.6\msun$ and $1.5 \msun$.

We slice the populations of the simulated and observed planets into the metal
poor ([M/H]~$< 0$) and the metal rich ([M/H]~$ > 0$) samples. In
fig. \ref{fig:sim_obs}, the top panel shows the PMFs for the simulated planets
inside the inner 5 AU. The blue, the red and the hashed green histograms show
the PMF for the total, the metal rich and the metal poor samples,
respectively.  The middle panel in fig \ref{fig:sim_obs} shows the observed
sample of planets (cf. fig. \ref{fig:pmf_mass}), also sliced by the
metallicity of the host.  No attempt to include errors or observational biases
in this panel was made.  For this reason, the observed PMF shown in the
  bottom panel massively under-estimates the number of low mass planets which
  are much harder to detect than high mass ones. However, the ratio of the
  metal poor to metal rich systems at a given planet mass is probably much
  more robust, and this is what we indent to compare with the simulations.  

The bottom panel of fig. \ref{fig:sim_obs} shows the fraction of stars that
are metal poor for three mass groups of planets. Qualitatively, similar trends
are seen in the simulations and in the observations: the Super-Earths are
(number) insensitive to the host star metallicity, medium mass giants are
positively correlated with the host star metallicity, and the most massive gas
giants are somewhat weakly correlated with the metallicity. The highest mass
bin in this picture is particularly interesting since, it appears to us, in CA
the highest mass giants should still correlate with the metallicity. These
planets are formed in CA by accreting more gas onto less massive giants, and
those do correlate with $z$. There is thus a potential to distinguish CA and
TD by the metallicity correlations for the highest mass gas giants.

As already mentioned, the observational sample for exoplanets used here
is unfortunately not uniformly selected and may have various selection
biases. \cite{AdibekyanEtal13} presented a study of metallicity correlations
for planets selected in a much more homogeneous way from the SWEET-Cat
database. Among a number of correlations, they find that most massive giants
in their sample (more massive than $4 \mj$) correlate with metallicity weaker
than less massive giants, although the statistical significance of this result
is not very robust due to a small number of planets in the high mass bin.

%
%
\section{Atmospheres on top of cores}\label{sec:atmo}

As described in \S \ref{sec:new_atmo}, in this paper we also calculate the
mass of gas gravitationally bound to the cores inside the pre-collapse
fragments. When the fragment is disrupted this atmosphere is assumed to
survive, remaining bound to the core. This is a reasonable assumption, since
the density of these atmospheres are typically higher than that of the main
body of the fragment by orders of magnitude \citep[e.g.,
  cf. ][]{NayakshinEtal14a}. The atmospheres could loose more mass after the
fragment disruption by photo-evaporation if the remnant migrates sufficiently
close to the star \citep[e.g.,][]{OwenWu13}; this post-disruption evaporation
is not included in our calculations at this stage.

\begin{figure}
\psfig{file=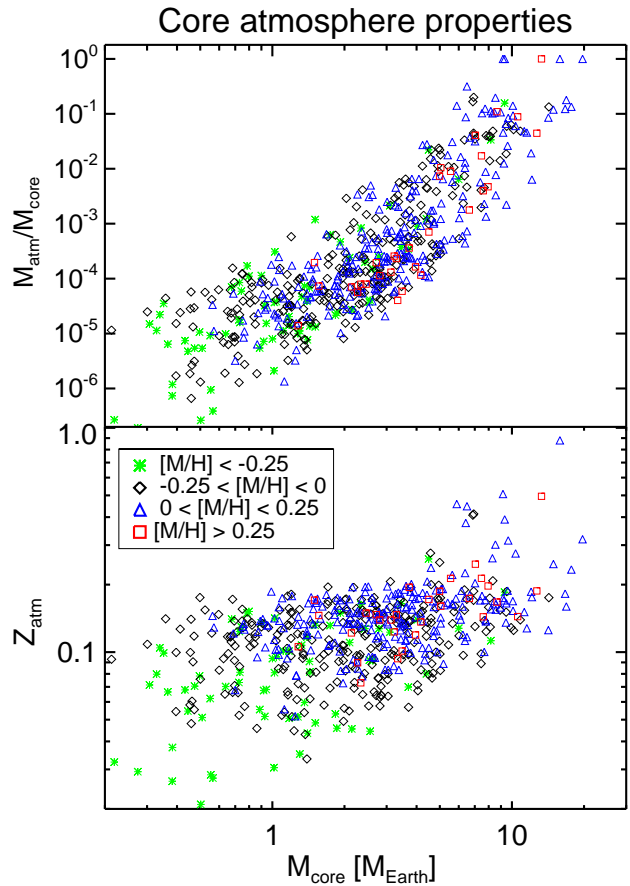,width=0.55\textwidth,angle=0}
\caption{Top: the ratio of the mass of the bound atmosphere to that of the
  core as a function of core mass after the fragment disruption. Bottom: the
  metallicity of the atmospheres shown in the top panel. The legend explains
  the meaning of the symbols.}
\label{fig:atmo}
 \end{figure}

The top panel of figure \ref{fig:atmo} shows the ratio of the atmosphere mass
to that of the core for the simulated planets in the inner 5 AU. Colours and
types of the symbols indicate the host star metallicity, as before. We see
that metallicity does not play a big role in establishing the mass of the
atmosphere at a given $M_{\rm core}$.  The metallicity is more important in
determining the core mass itself, as we saw in \S \ref{sec:SE}. One notices
that metal rich systems (blue and red) appear more often at $M_{\rm core}
\simgt 5 \mearth$ part of the figure, whereas the metal poor systems typically
have smaller mass cores.

Notice that there is a large spread in the relative atmosphere mass at a given
$M_{\rm core}$. This spread is not a function of metallicity. The physical
reason for the spread is that cores of a given mass can be born inside
fragments of different masses and/or different evolution histories. These
fragments have different central gas pressure and densities which explains why
same mass cores may have different atmosphere masses as measured at the
fragment disruption \citep[see also][]{NayakshinEtal14a}. In addition, we
parameterise the core's luminosity by introducing the Kelvin-Helmholtz time for
the core, $t_{\rm kh}$, which is a Monte Carlo parameter varied between $10^5$
and $10^7$ years (see Table 1 and \S \ref{sec:new_atmo}). Due to this, the
luminosity of the cores vary, and that also imprints onto the scatter of the
atmosphere's mass at a given $M_{\rm core}$.

With all the scatter, it is nonetheless clear that cores less massive than
$\sim 5\mearth$ do not possess atmospheres more massive than a few percent of
the core, whereas cores more massive than that may have atmospheres of mass
comparable to that of the core. A few of these formally had atmospheres more
massive than the core, which we capped at $M_{\rm atm} = M_{\rm core}$ exactly
as explained in \S \ref{sec:new_atmo}. Physically in these cases we expect
that a tidal disruption of the fragment will result in a partial disruption of
the fragment only as there can be more mass in the collapsed atmosphere around
the core than the core mass itself. At the same time, we note that in
fig. \ref{fig:atmo} there are also quite massive core, $M_{\rm core}\simgt 10
\mearth$ that have very small, $\sim 1$\%, atmospheres.

The bottom panel of fig. \ref{fig:atmo} shows the metallicity of the
atmospheres of the cores, $Z_{\rm atm}$. All of these values are strongly
super-Solar ($Z_\odot = 0.015$). This is to be expected as gas fragments
accrete pebbles in our model, so their abundances are super-Solar (cf. \S
\ref{sec:z_inside} on this for more detail). In addition to that, grains do
sediment to the centre of the fragment and so the metal abundance in the
atmosphere is usually even higher than the mean fragment metallicity (e.g.,
figure 9 in paper I shows that the metallicity is the highest in the centre of
a planet).

%
%
\section{Rock dominated composition of cores}\label{sec:core_comp}

\begin{figure}
\psfig{file=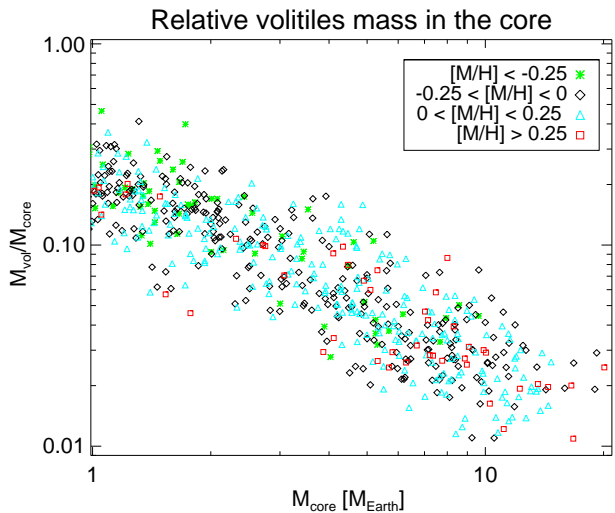,width=0.55\textwidth,angle=0}
 \caption{The ratio of the core mass in the volatiles (water ice and CHON) to
   the total mass of the core as a function of the core mass. Note that high
   mass cores are strongly dominated by rocks in our model, in contrast to
   CA.}
\label{fig:cores}
 \end{figure}

Figure \ref{fig:cores} shows the ratio of the volatile (water and CHON) mass
in the core to the total mass of the core, again for the four different
metallicity groups. We see that with a few exceptions at the low core mass
end, the cores are always dominated by the rocks rather than ``ices''. This
is especially true at the high core mass end, where CHON contribute a few
\% of the mass only.

This result is not particularly surprising in the light of previous studies of
grain sedimentation inside the pre-collapse fragments. \cite{HS08} showed that
water ice grains are unlikely to sediment into the cores of pre-collapse gas
fragments of mass higher than about $1 \mj$ because the fragments are too hot
for the icy grains to exist even at early times. \cite{HelledEtal08} then
showed that organics (modelled as CHON) grains have somewhat better chances of
sedimenting into the cores since their vaporisation temperatures are $\sim
350$ to 400~K, depending on gas pressure, whereas water ice vaporises above
$T\sim 150-200$~K. Rocks and Fe (counted together as one species in our paper)
are thus the main components of which the cores are made in the TD model
\citep[see also][]{ForganRice13b}.

There remain {\em quantitative} uncertainties in the composition of the cores.
We here chose a rather low initial temperature for the fragments, so that
water ice can sediment initially in gas fragments less massive than a few
Jupiter masses. This is why there are some volatile grains in the relatively
low mass cores in fig. \ref{fig:cores}. However, high mass cores require a
long, $\simgt$ a few $10^4$ years, assembly time. The fragments containing
those heat up to temperatures of hundreds of K during this time, so that
neither CHON or water ice grains can sediment.

This prediction of the model is important as it is very different from CA
model in which massive cores have the best chance of growing beyond the ice
line where water ice can sediment, so that they are dominated by ices.

\section{Metal overabundance of gas giants composition}\label{sec:z_inside}

Gas giant planets, and also their less massive cousins, are over-abundant in
elements heavier than H/He compared with their host stars. This is well known
for the Solar System planets \citep[e.g.,][]{GuillotEtal04} and is now also
robustly established for exoplanets at close separations
\citep{MillerFortney11}. This important fact is usually used
\citep[e.g.,][]{HelledEtal13a} to argue that the CA model is a clear favourite
to explain giant planets formation because planets made by CA contain heavy
cores which makes the planets more compact (as required by the observations)
than they would be if they had their host star metallicities and did not
contain massive cores. In contrast, GI planets may be enriched in metals by
accretion of grains or planetesimals early on \citep{BoleyDurisen10}, but
until recently it was thought that this enrichment is not fundamental to the
survival of GI planets. It would thus seem that GI planets may have a range of
metallicities -- from sub-host values to many times the metallicity of the host
\citep{BoleyEtal11a}.

However, as argued in \cite{Nayakshin15a}, pebble accretion is the process
that makes relatively low mass (a few $\mj$) GI fragments survival
possible. Radiative cooling is too inefficient, and it is pebble accretion
that allows these fragments to contract into the much denser ``hot start'' GI
planet configurations so that they avoid tidally disruptions in the inner
disc. Therefore, we may expect that the metallicity of TD giant planets is
elevated compared to their host stars.

\begin{figure}
\psfig{file=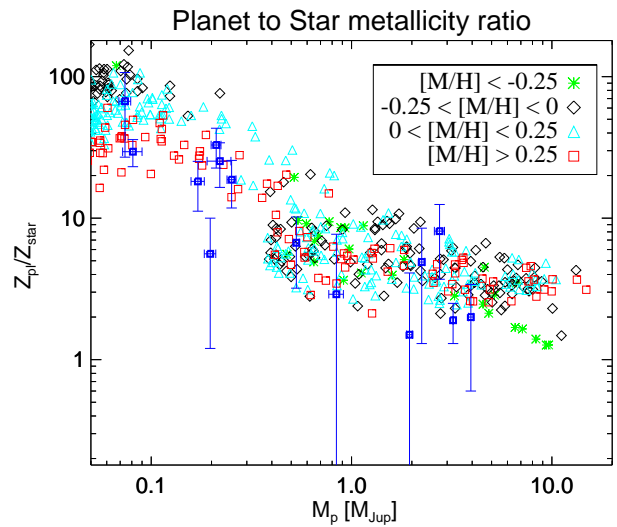,width=0.55\textwidth,angle=0}
 \caption{Ratio of metal abundance of the giant planets to that of their host
   star. The symbols with the error bars are models by Miller \& Fortney 2011
   based on observations of close-in but not strongly irradiated gas giant
   planets.}
\label{fig:ZvsM}
 \end{figure}

Figure \ref{fig:ZvsM} shows the ratio of the metallicity of the simulated
planets in the inner 5 AU, $Z_{\rm pl}$, to the metallicity of the host star
as a function of the planet's mass. As before, the population is broken into
four bins by the host star metallicity, [M/H], as indicated in the legend. In
addition to that, we used the ``mixed model'' in Table 1 of
\cite{MillerFortney11} to show their best estimates for such a ratio for their
sample of observed giant planets. These models are shown with the blue symbols
with the error bars in fig. \ref{fig:ZvsM}.

Fig. \ref{fig:ZvsM} indicates that TD giant planets are indeed over-abundant
in metals compared to their host stars. In addition, there is an
anti-correlation between the ``over-metallicity'' ratio, $Z_{\rm pl}/Z_{\rm star}$, and
the mass of the planet. The origin of this anti-correlation is slightly
different for planets that did not experience tidal disruptions, $M_p \simgt
0.33 \mj$ here, and those that did (the less massive ones). 

For the partially disrupted planets, the correlation is physically similar to
that of CA gas giant planets. The metal content of these planets is actually
dominated by the core's mass, which is $M_{\rm core}\sim 10\mearth$. The
atmosphere, while also over-abundant in metals as we saw in
fig. \ref{fig:atmo}, contains a minority of the metals mass. Since $M_{\rm
  core} \sim$~const, and is independent of the total planet mass, $Z_{\rm pl}
\sim M_p^{-1}$ is this regime.

For heavier planets that did not go through a tidal disruption, the
correlation occurs because (a) higher mass planets are hotter to begin with
\citep[e.g.,][]{HS08,Nayakshin10b} and (b) collapse faster as radiative
cooling contributes more to their contraction \citep{Nayakshin15a}. Therefore,
more massive gas fragments require a smaller fractional mass increase in
pebbles to reach the central temperature of $\sim 2000$~K and collapse.

Note that we do not take into account gas accretion onto the planets in this
paper in either pre- or post-collapse evolution. If this assumption was
relaxed, post-collapse planets could gain mass by accretion of gas with
metallicity close to that of the host star. This would therefore dilute their
metallicity at a given total planet mass, but, importantly, would not change
the downward trend that we see in fig.  \ref{fig:ZvsM}.

Our simulation results are qualitatively consistent with the models of
\cite{MillerFortney11}, although we did not fine tune any of the parameters to
specifically address this issue. We finish this section reiterating what we
said in \S \ref{sec:td_valley}: the present day bulk structure of the planets
may not be uniquely indicative of the route that these planets formed. We
believe TD is as promising as CA in explaining the bulk composition of Solar
and extra solar planets.

%
%
\section{Which Fragments make which planets?}\label{sec:which}

\begin{figure}
\psfig{file=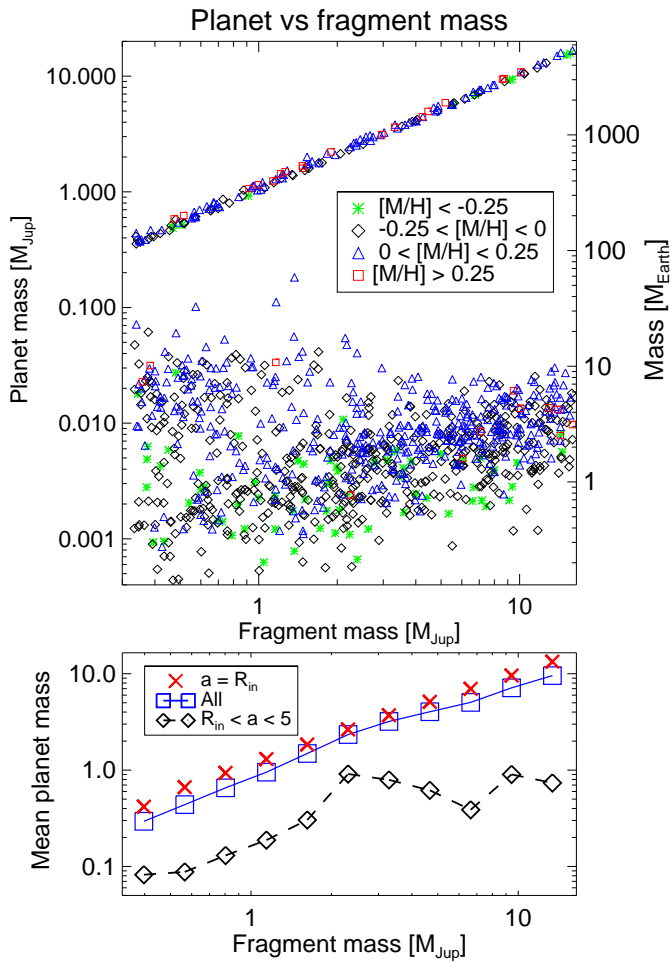,width=0.5\textwidth,angle=0}
 \caption{Top: Final mass of the planet versus the initial mass of the
   fragment. The meanings of the symbols is the same as in
   fig. \ref{fig:atmo}. Bottom: Mean mass of the planets more massive than $5
   \mearth$ versus initial fragment mass for three regions in the host-planet
   separations.}
\label{fig:Mfinal}
 \end{figure}

One may wonder how the value of the initial fragment mass maps into the kind
of planets that emerge in the end. To investigate that, the top panel of
figure \ref{fig:Mfinal} shows the final mass of the planet versus the initial
fragment mass, $M_f$, for a uniformly randomly chosen (7\% of the total)
subsample of all the planets in the simulations. The colours reflect the
metallicity of the system, as before. The nearly straight line in the upper
part of the panel are the fragments that avoided tidal disruption and
collapsed into giant gas planets. Mass of these planets is close to the
initial fragment mass, save for the addition of metals by pebble accretion;
hence the (almost) one to one correspondence between the final and the initial
planet masses.

The planets below that line are planets that went through a total or a partial
gas envelope disruption. These are mainly rocky core planets with only a few
systems that are dominated by gas, as we saw earlier.

This figure demonstrates that the mass of the core made within the fragment is
not a unique function of the fragment's mass. In fact, the most massive cores
are found in the least massive fragments, those below $M_f \sim 2 \mj$. This
result is not new. \cite{HelledEtal08,HS08} showed that the more massive the
fragment is, the less massive is the core made inside it during the
pre-collapse phase. This is because the central temperature of pre-collapse
fragment is believed to increase with its mass even at the initial stage due
to adiabatic compression being stronger for more massive fragments
\citep{Masunaga00,Nayakshin10a}. Massive fragments also cool radiatively more
rapidly, cutting the available time for the core's growth. The exact value of
the initial temperature, $T_0$, is however difficult to calculate from first
principles, especially in the case of the disc. In this paper we assumed that
$T_0 \propto M_f^{1/2}$. A stronger dependence would make even smaller cores
at high $M_f$ than the top panel of figure \ref{fig:Mfinal} shows.

The bottom panel of fig. \ref{fig:Mfinal} presents the mean mass of the planet
as a function of the initial fragment mass. This is calculated neglecting the
least massive planets, $M_p < 5 \mearth$, assuming that these are unlikely to
be detected due to selection bias. Further, three spatial regions are picked
for this panel: the planets on the very edge of the disc (red crosses), planet
found anywhere in the disc (blue squares) and planets found between the inner
edge and 5 AU (black diamonds). 

In general, we see that the mean mass of the resulting planet correlates
positively with the mass of the initial fragment. This result is of no clear
importance if the spectrum of initial fragment masses is same for all the
systems studied -- as assumed in this work. This may not necessarily be so,
and further work on the gravitational fragmentation of discs is needed to
study this issue. One rather likely outcome, however, is that the mass of the
fragment would increase with the mass of the host star. If this is the case
then the  bottom panel of fig. \ref{fig:Mfinal} predicts that the mean mass of
a planet would increase with the mass of the host star.

\begin{figure}
\psfig{file=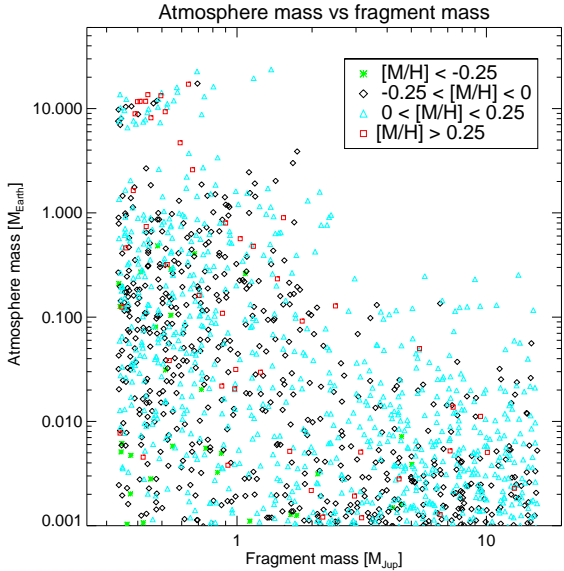,width=0.45\textwidth,angle=0}
 \caption{Mass of the core's atmosphere versus the fragment's initial
   mass. Note that all fragments, whether disrupted or not, are plotted. For
   non-disrupted fragments the atmosphere simply becomes the part of the much
   larger gaseous envelope and is not distinguishable from it.}
\label{fig:Matmo}
 \end{figure}

The most massive cores, $M_{\rm core} \sim 10 \mearth$ are made in relatively
low mass fragments, $M \simlt 2 \mj$, as emphasised before. We can hence
expect a similar result for massive atmospheres {\em bound} tho the
core. Fig. \ref{fig:Matmo} indeed shows that cores with massive atmospheres
are only made in low mass fragments. This result is interesting in that it
predicts that in environments in which only massive fragments are born in the
disc, there is very little chance of finding any planet with mass intermediate
between a few Earth masses and a few Jupiter masses. Fig. \ref{fig:Matmo}
shows the mass of the atmosphere irrespectively of whether the fragment is
disrupted, in which case the atmosphere remains bound to the core, or
collapses, in which case the ``atmosphere'' simply becomes the part of the
much more massive gaseous envelope of the planet.

%
%

\section{Radial distribution of planets}\label{sec:radial}

We finally discuss the distribution of the simulated planets over the planet
-- host separations. This is one of the most interesting
observables. Unfortunately it is also one of the most uncertain results of our
model at this point. While many of the planet properties studied in previous
sections, including the metallicity correlations, depend mainly on the planet
formation processes of TD (fragment formation, contraction, core growth, tidal
disruption, etc.), the radial distribution of planets depends at least as
sensitively on the protoplanetary disc evolution model accepted here. Section
\ref{sec:uncertain} shows why this statement is true for our simulated models,
but it is also bound to be true for any planet formation theory
\citep[e.g.,][]{AA09}. These uncertainties go beyond planet formation theory
and cannot be resolved until evolution of protoplanetary discs is understood
in sufficient detail. We believe that real protoplanetary discs are far more
complex than the simple model accepted here or in any other published
population synthesis models
\citep[e.g.,][]{IdaLin04a,IdaLin04b,IdaLin08,MordasiniEtal09a,
  MordasiniEtal12}.  This view is motivated by the increasingly popular ideas
of episodic accretion on young stars \citep[e.g.,][]{DunhamVorobyov12} that
may have direct connections to GI model \citep{BoleyEtal10,NayakshinLodato12},
and observational surprises in the population of "transition discs" that were
expected to be the link between protoplanetary and debris discs (see the end
of the Discussion section).

\begin{figure}
\psfig{file=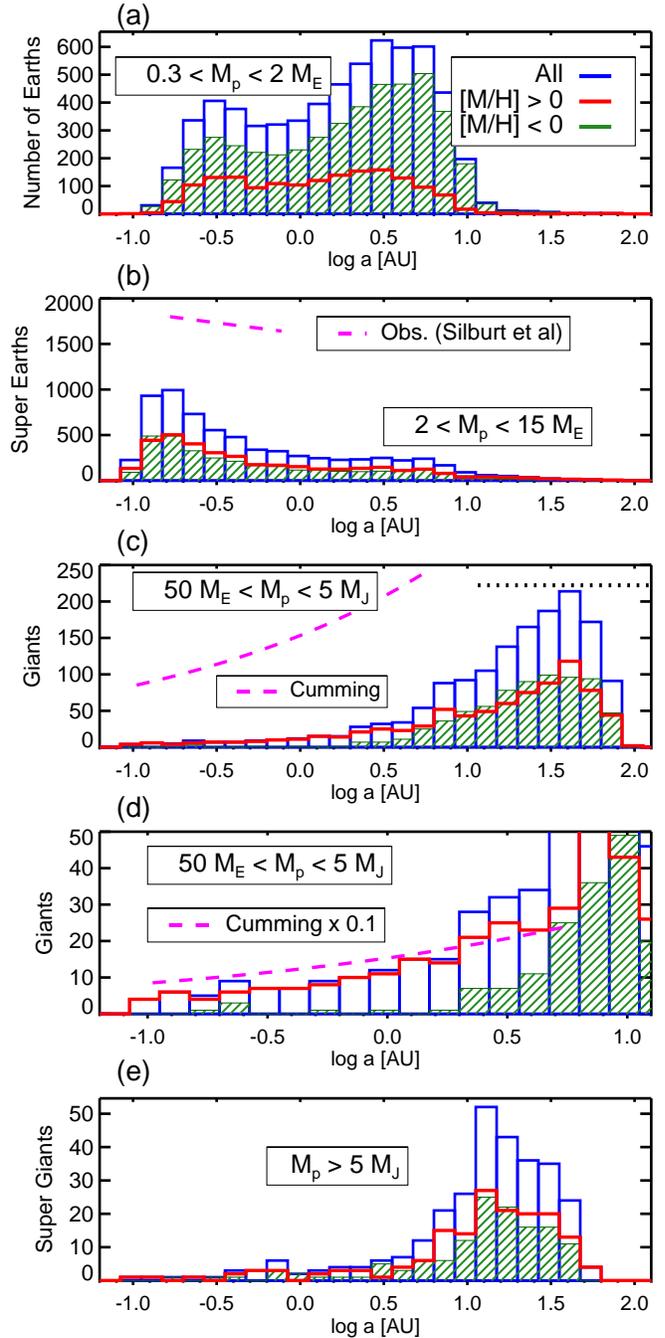,width=0.5\textwidth,angle=0}
 \caption{The histogram of planet final separations for four groups of
   planets: Earths, Super Earths, Giants and Super Giants, from top to
   bottom. The colours show all, metal rich and metal poor populations, as
   indicated in the legend on the top panel.}
 \label{fig:separation}
 \end{figure}

With this caveat in mind, fig. \ref{fig:separation} shows how the planet final
separations are distributed for four different groups of planets. The top
panel (a) shows ``Earths'', planets with mass $0.3\mearth < M_p < 2
\mearth$. Panel (b) shows Super Earths defined to be planets with mass
$2\mearth < M_p < 15 \mearth$. The third and the fourth panels show ``giants''
with mass between $50 \mearth$ and $5 \mj$. Most massive gas giants, $M_p > 5
\mj$, are shown in panel (d). The histograms are further broken into the metal
rich (red), the metal poor (green shaded) and the total populations (blue
colour). We do not show in fig. \ref{fig:separation} the ``assimilated'' group
of planets, that is those that migrated all the way to $a = R_{\rm in}$, since
we cannot constrain their further fate here.

\subsection{Low mass cores}\label{sec:coresR}

Panel (a) of fig. \ref{fig:separation} shows that Earth-like core-dominated
planets are most abundant at $a \sim$ a few AU separations. Physically, this
is because fragments migrating from the outer disc are disrupted most
frequently in this region \citep[this can be deduced already from simple
  analytical estimates, see, e.g.,][]{Nayakshin10c}, and hence this is where
these cores are deposited by the disruptions. Low mass cores do not migrate
inward significantly, so they form the dominant peak in the same spatial
region. One also observes that the histogram for the low mass rocky planets is
mainly composed of cores born at low metallicity environments, which is to be
expected given the results of \S \ref{sec:Zcor}.

One may also note that the metal rich population of low mass cores forms a
flatter distribution of host -- planet separations which is shifted a little
closer to the star. This can be attributed to two effects. Firstly, the cores
born in metal rich environments are born in more compact gas fragments. These
fragments are disrupted closer to the host star. Secondly, cores around metal
rich stars are more massive on average that those in the low metallicity
discs, and so they also migrate further in.

\subsection{Super Earths}\label{sec:seR}

Panel (b) in fig. \ref{fig:separation} indicates that super Earth planets are
shifted towards smaller separations with respect to the low mass cores. This
is due to a higher inward migration speed of the super Earths which brings
most of them into the inner region. For this reason the super Earths dominate
the mass function at small separations, as was already seen in
fig. \ref{fig:pmf_cores}. The metal rich and the metal poor populations of the
super Earths are quite similar, which again shows that there is no metallicity
correlation for these planets.

\cite{SilburtEtal15} reconstructed the frequency of occurrence of {\em Kepler}
detected planets in the size range $1-4 R_\oplus$ in the orbital period range
from $\sim 10$ to 200 days. The dashed purple line in panel (b) in
fig. \ref{fig:separation} shows their result, which should approximately map
to the mass range we use for this panel. We see that there is a discrepancy of
about a factor of three between the observations and our one fragment
simulations. Another way of putting this is to say that if different fragments
did not interfere with each other (which is most likely not true), then we
would need $\sim 3$ fragments per star to reproduce the observed abundance of
super Earths inside the inner $\sim 1$~AU.

We note that the outer region of $\sim$ tens of AU contains very few cores in
our simulations (see also fig. \ref{fig:scatter}). This outcome stems from
tidal disruptions of fragments becoming possible only in the inner $\simlt
10$~AU region. This result does depend on assumptions and parameters of the
model, such as the fragment migration speed, pebble abundance in the disc, the
initial disc mass, etc. For example, simulations in paper II did produce
massive cores in the outer tens of AU region (cf. fig. 4 in that paper).

\subsection{Gas giants}\label{sec:giantsR}

Panels (c) and (d) of fig. \ref{fig:separation} show the same simulated
population of planets with mass between $50 \mearth$ and $5 \mj$, except panel
(d) is a zoom-in on the inner $10$~AU part of the histogram.  These panels
show that very few of the gas giant planets actually end up in the region
between $0.1 < a\simlt 5$~AU. The most common outcome (cf. also
fig. \ref{fig:outcomes}b) for the initial fragments is to migrate inward
rapidly and be destroyed or to collapse but end up on the inner edge of the
disc at $a=0.09$~AU. The next most common outcome is that some fragments are
"stranded" in the outer region by the disc disappearing more rapidly than the
fragments can migrate.

The fraction of the simulated gas giants in the inner disc appears too low
compared with the observations. \cite{CummingEtal08} show that the occurrence
rate of gas giant planets with period less than 5.5 years and in the mass rate
$0.3 \mj$ to $10 \mj$ is $\approx 10$\%. The frequency of the observed giants
around FGK stars from \cite{CummingEtal08} is shown in panel (c) of
\ref{fig:separation} with the purple dashed line.

\cite{BillerEtal13} places a model-dependent upper limit of $\sim 10$\% on the
frequency of $1-20\mj$ directly imaged mass planets at separations $10 -
150$~AU. This shows that systems such as HR 8799 are very rare. Assuming a
flat in $\log a$ distribution for such planets, we plot the
\cite{BillerEtal13} result as a black dotted line (recalling that we have
20,000 host stars in our simulations).

Panel (c) shows clearly that while our model is marginally consistent with the
upper limits on the directly imaged planets, the number of planets in the
inner $\sim 10$ AU is very much smaller than observed. Physically, this is due
to a very rapid inward migration of the giant planets through this region on
the way to $a = R_{\rm in}$, due to which it is quite unlikely that a planet
would stop there at the time when the disc is dissipated away.

The zoom-in on the inner part of panel (c) is shown in panel (d). It shows
that the fraction of medium mass giant planets in our sample (which is about
1.0 \%) is almost exactly 10 times lower than the observed gas giant fraction
(the purple dashed curve in panel [d] is same as in panel [c] but is
multiplied by the factor of 0.1).

Evidently, the simulations could be reconciled with observations by either a
different disc model which would have the planets migrate slower through the
inner region, or by having $\sim 10$ fragments per host star initially and
removing the ``excess'' far away giants by some process. Close planet-planet
scattering during the formation phase \citep{VB13b} and stellar interactions
may remove some of the far-out giants \citep{DaviesMEtal14}, but it seems
doubtful that these processes would be effective in removing as much as 90\%
of the original population.

One feature in the radial distribution of gas giant exoplanets not
  reproduced by power-law fits such as that by \cite{CummingEtal08} is the
  sharp ``pile-up'' of planets at $a\simgt 1$~AU \cite{WrightEtal09}. Models
  by \cite{AlexanderPascucci12} show that photo-evaporation of the disc may
  produce a similarly strong pile-up of planets at about the right
  place. Physically, when the disc is dissipated by photo-evaporation, the gas
  is removed most rapidly from the region with $R\sim$ a few AU, which then
  implies that the planets migrate through this region slower than one would
  expect from a non photo-evaporating disc. This then forms a spike on the
  final semi-major distribution of gas giants. Note that this physics is
  independent of how the gas giant planets are formed, although there is a
  strong dependence on the timing of the planet ``injection'' into the inner
  few AU of the disc.

  Although our models include disc photo-evaporation, we do not find such a
  strong pile-up of gas giants in our calculations. However, further tests, to
  be presented elsewhere, show that pile-up formation is a strong function of
  the disc viscosity parameter $\alpha_0$. In this paper we kept it at a fixed
  ``reasonable'' value $\alpha_0 = 0.005$ (cf. \S 2.2.1). Simulations with a
  broad range of $\alpha_0$ actually do result in a cliff-like decrease of gas
  giants inward of $a\sim 1$~AU. We plan to analyse and present this issue in
  future work.

Finally, panel (e) of fig. \ref{fig:separation} shows the radial distribution
of planets more massive than 5 $\mj$. As for less massive planets, most of
these are located at large separation. There is no strong metallicity
correlation either in the inner or the outer disc for these planets.

\subsection{Disc evolution impact on planet yields}\label{sec:uncertain}

As can be gathered from \S \ref{sec:disc_evol} and \S \ref{sec:MonteCarlo},
even with a fixed radial structure of our initial discs and a small -- only a
factor of two -- range in the initial total disc mass, the time scale for the
protoplanetary disc dissipation varies widely between the runs in line with
the observed range in the disc lifetimes (fig. 1). It may be expected that the
duration of the disc lifetime has a significant impact on how the fragments
evolve. To explore this issue, we define an "evaporation time", which is an
estimate of the time scale on which the protoplanetary disc would be removed
by photo-evaporation,
\begin{equation}
t_{\rm ev} = {M_{\rm d}\over \dot M_{\rm ev}}\;,
\label{tevap}
\end{equation}
where $M_{\rm d}$ and $\dot M_{\rm ev}$ are the initial disc mass and the
total photo-evaporation rate, respectively. We then define the planet yield as
the frequency of a given type of planet formation for four planet sub-samples:
hot giants at the inner disc edge, e.g., all planets more massive than $50
\mearth$ with the final planet-host separation $a=R_{\rm in}$; hot and cold
giants -- same mass range but $R_{\rm in} < a < 5$~AU and $a > 10$~AU,
respectively; and hot sub-giants, $R_{\rm in} < a < 5$~AU and $M_p <
25\mearth$. The exact dividing masses are not important for what follows.

\begin{figure}
\psfig{file=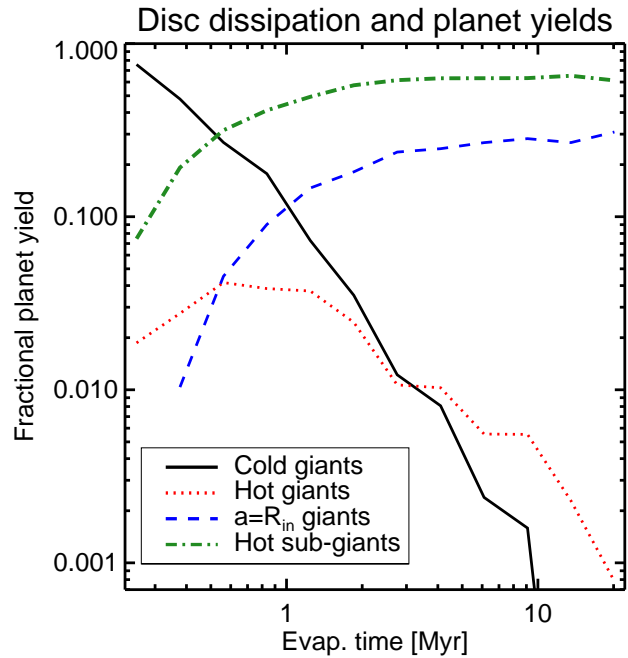,width=0.5\textwidth,angle=0}
 \caption{The frequency of cold and hot giant and hot sub-giant planet
   formation in our full simulated planet sample versus the estimated disc
   photo-evaporation time. The longer the disc exists, the smaller the chance
   of a cold giant planet formation, and the higher the chance that either a
   super-earth planet forms or that the gas giant is pushed all the way into
   the star ($a=R_{\rm in}$).}
 \label{fig:yield}
 \end{figure}

Figure \ref{fig:yield} shows how the planet yields vary with the evaporation
time. The meaning of the trends are physically clear. The faster the disc is
evaporated, the smaller is the chance of a gas giant planet being "stranded"
in the outer disc. In the limit of no photo-evaporation, $t_{\rm ev}
\rightarrow \infty$, our discs always manage to push the gas giants into
the inner disc. Discs that are removed very quickly (short $t_{\rm ev}$), on
the other hand, disappear sooner than they are able to push the planet close
to the star. For shortest $t_{\rm ev}$ bins, most of the initial fragments end
up as cold giants.

For the hot giant planets the trend is not monotonic. At short $t_{\rm ev}$,
the longer the disc is present, the more of the cold giants end up closer to
the star, hence the yield of the hot gas giants increases as $t_{\rm ev}$
increases. However, this upward trend levels off and then turns into a
decreasing one at longer $t_{\rm ev}$ since then the hot giants are more
likely to be pushed all the way to $a=R_{\rm in}$, which is clear from the
dashed blue curve, or be disrupted and become a sub-giant planet (the green
dot-dash curve).

Figure \ref{fig:yield} demonstrates that the outcome of planet formation
process in the context of TD hypothesis for planet formation strongly depends
on the disc model accepted, and this dependence is quite varied and may be non
monotonic for different types of planets. While more experiments with
different disc models are needed, it appears likely to us that the fraction of
the cold giants formed in our models may be reduced by reducing the maximum
photo-evaporation rate in our models (which is indeed very high, e.g., $
{\rm max}[\dot M_{\rm ev}] = 3 \times 10^{-7} \msun$~yr$^{-1}$). We therefore do not consider
the mismatch in the radial distribution of our simulated planets as an
unsurmountable challenge to the TD model.

%
%
%

\section{Discussion}\label{sec:discussion}

This paper presented a synthesis of 20,000 planet formation experiments in the
context of a relatively young planet formation theory called Tidal
Downsizing. TD planet formation process begins as in GI model, with
  formation of a massive gas fragment in the outer self-gravitating gas
  disc. Instead of stopping there artificially, TD continues with (a)
  migration of the fragment in, as found in a dozen independent numerical
  simulations; (b) grain sedimentation forming the massive solid core in the
  centre of the fragment; (c) either collapse of the fragment, which results
  in a very young gas giant planet, or tidal disruption of the fragment once
  it migrated too close to the host star, which forms a core-dominated
  planet. TD model is thus GI plus modern physics, and is an attempt to
  rectify a certain injustice that GI model suffered while attention of the
  modellers in the last decades was focused on CA almost entirely.

The most important result following from our calculations is that many of the
observed properties of the Solar System planets and exoplanets that were
previously claimed to support CA theory uniquely are naturally reproduced by
the TD theory as well. In particular, we find that 
\begin{enumerate}

\item giant planets do contain massive cores (\S \ref{sec:which}). Their mass
  is dependent on the dust growth and sedimentation physics and the initial
  conditions for the fragments. At this point we cannot rule out nearly zero
  core masses for some giant planets, especially for those more massive than a
  few $\mj$, since they contract rapidly and may vaporise their grains too
  soon \citep[cf. also][]{HS08}. The most massive cores formed in the
  simulations are $\sim 10-20\mearth$ (fig. \ref{fig:pmf_cores});

\item gaseous envelopes of gas giants are strongly enriched in metals compared
  with their host stars (especially in organics and water since rocks are more
  efficient in sedimenting into the core), see \S \ref{sec:core_comp};

\item the over-metallicity of gas giants decreases as the planet mass increases, as
  observed for the Solar and some giant exoplanets (\S \ref{sec:z_inside});

\item the model is able to produce copious core-dominated planets with mass
  from sub-Earth up to $\sim 20 \mearth$. These planets are the most frequent
  outcome of TD planet formation hypothesis, but less so at high metallicities
  when giant planet survival becomes much more likely
  (cf. fig. \ref{fig:outcomes}b);

\item a positive correlation of frequency of appearance of gas giant planets
  with metallicity of the host star for moderately massive giants in the inner
  $\sim 5$~AU region (cf. \S\S \ref{sec:Zcor} and \ref{sec:SE}).

\item Same planets do not show a metallicity preference at large separations
  ($a > 5-10$~AU) from their host stars (\S \ref{sec:cold_giants});

\item The most massive giant planets do NOT correlate positively with
  metallicity (cf. \S \ref{sec:BDs}). Also, extending the results of \S
  \ref{sec:z_inside} into the higher mass regime, BDs and low mass stellar
  companions formed by GI of their discs (and then having accreted more gas
  from it) have no physical reason to be over-abundant in metals compared with
  their host stars;

\item Core-dominated planets show a complicated pattern of occurrence versus
  the host star metallicity dependence. While the average core mass increases
  with metallicity, the number of the cores formed in the disc does not (\S
  \ref{sec:SE}). The latter is directly connected with the fact that more
  initial gas fragments survive at high metallicities, which implies fewer
  tidal disruptions, and hence fewer core-dominated planets. Sub-Neptune mass
  planets thus correlate in mass but not in numbers with the host star
  metallicity. We emphasise that this prediction stems from the basic
  mechanics of assembly of different types of planets in TD, e.g., more tidal
  disruptions equals less gas giants but more core-dominated planets. It is
  hence very robust, unlike the CA result that depends on detail of type I
  migration prescriptions \citep{MordasiniEtal12};

\item the composition of simulated cores is dominated by rock and Fe, not
  ices, especially at high core mass end (\S \ref{sec:core_comp}). This
  prediction is significantly different from CA model. To remind the reader,
  core compositions are currently far from a settled issue even in the Solar
  System: while Uranus and Neptune are often referred to as "ice giants",
  there is no direct evidence that their bulk composition contains dominant
  amounts of ice. Modelling of {\em Voyager} and other data for these planets
  with a wide range of equations of state shows that compositions of these
  planets {\em could be} dominated by pure rock not ice
  \citep[see][]{HelledEtal11};

\item due to a variety of fragment evolution histories, there is no one-to-one
  relation between the atmosphere mass and the core mass for the simulated
  planets. However, in general, the masses of the atmospheres of
  core-dominated planets are a tiny fraction of the total planet mass for
  cores less massive than $M_{\rm core}< 5 \mearth$, but become comparable to
  the core's mass for $M_{\rm core}\simgt 5-15 \mearth$ (\S
  \ref{sec:atmo}). The atmospheres are also metal-rich (fig. \ref{fig:atmo});

\item the planet mass function is dominated by the sub-Neptune mass planets,
  and contains a pronounced ``tidal disruption desert'' between $M_p \sim
  20\mearth$ and $M_p \sim 100\mearth$ (cf. \S \ref{sec:td_valley});

\item the planet mass function does not run away towards low mass ($M_{\rm core}
  \simlt 1 \mearth$) cores, unlike CA mass function (cf. \S
  \ref{sec:no_pool}). 

\item The PMF is smooth and does not divide on
  ``rock-dominated'' and ``ice-dominated'' cores (\S \ref{sec:no_minima}).

\item The PMF has a rollover at mass $\sim 10 \mearth$ since it appears
  difficult to make more massive cores (\S \ref{sec:massive_cores}).

\end{enumerate}

Point (vii) may have significant implications beyond planet formation
theories. It is usually said that formation scenarios of giant planets must be
physically different from that of brown dwarfs and low mass stellar companions
\citep[e.g., see references in \S 2.1 of][]{WF14}, mainly because the
metallicity correlations of these objects are different. While giant planet
frequency of detection is positively correlated with metallicity of the host,
for BDs such a correlation does not seem to exist conclusively, and for the
low mass stellar companions it is the low metallicity hosts that are more
likely to host the companion \citep{RaghavanEtal10}. We find that there may be
physical reasons within the TD formation scenario that explain the divergent
metallicity correlations for these different objects. In particular, if
formation of planets starts with the birth of a $M_f \simlt $ a few $\mj$ gas
fragment in a gravitationally unstable disc, and it does not grow more massive
by gas accretion, then pebble accretion favours the survival of close-in
exoplanets in high metallicity environments.  If formation of BDs and low mass
stellar companions starts with the birth of a more massive seed, e.g., $M_f
\simgt 10 \mj$, then survival of these {\em by radiative cooling} is favoured
at low metallicities \citep[see also][]{HB11}. We found this low-opacity mode
of collapse to become important already at $\sim 10\mj$ in \S \ref{sec:BDs},
and it is clear that for higher mass fragments radiative cooling will dominate
over the pebble accretion collapse even more.

If stellar mass secondaries can indeed grow from low mass $M_f \simgt 10\mj$
seeds born in the early embedded phases of the star evolution, then there are
further interesting implications stemming from TD model for the formation of
planets in stellar binary systems.  {\em Kepler} observations show that
planets orbiting binary stars may be as common as $\sim 1-10$\%
\citep{WelshEtal12,SchwambEtal13}. This is surprising in the context of
CA. Planetesimal velocity dispersions are whipped by the secondary in stellar
binary systems \citep[e.g.,][]{PaardekooperEtal08}. Planetesimals are expected
to collide in such systems at velocities from tens of m/s to $\sim 1$ km/s and
fragment rather than grow \citep{PaardekooperEtal12,LinesEtal14}, making it
challenging to explain the observed systems. In contrast, formation of these
systems does not appear to strain any physical limits in TD but it does
require the following chain of events to occur. In particular, suppose that
several gas fragments are born in the outer disc, and one of them is much more
massive than the rest and grows by gas accretion into a stellar secondary. It
does not obviously precludes the other low mass gas fragments from maturing
into planets although quantitatively their evolution is probably different
from the one in a disc around a single star. The planet-mass fragments can be
born inside or outside of the orbit of the more massive fragment.  Formation
of either circum-primary or circum-binary planets can then be achieved by
shrinking the separations of the planet and/or the secondary due to the
gravitational torques of the disc, assuming the disc is massive enough or is
continued to be supplied with more matter from the outer envelope.

One feature of the model which is at odds with the observations is the radial
distribution of the giant planets, with the ratio of cold (planet-host
separation $a > 10$ AU) giants to warm giants ($a < 5$ AU) being larger than
observational limits by a factor of 10 or more. Our simulations also did
  not reproduce the period valley \citep{WrightEtal09} in that the observed
  one is between $0.1 \simlt a \simlt 1$~AU, whereas ours is from $0.1\simlt a
  \simlt 5$~AU.

The number of far-out gas
giants can be reduced if some of them self-destruct due to an overly luminous
core, as in the proposed formation route for Neptune and Uranus by
\cite{HW75}, see also \cite{NayakshinCha12}. In addition, the radial
distribution of giants is also very sensitive to the disc evolution model,
e.g., the very poorly known disc viscosity parameter $\alpha$, the locations
of fragments birth, and also the very end phases of the disc dispersal (cf. \S
\ref{sec:uncertain}). We believe that the current paradigm of protoplanetary
disc evolution, which is the base of our disc model here, is unlikely to be
entirely correct. This paradigm, developed with CA model in mind, all but
discounts the importance of massive gas planets and brown dwarfs for the disc
evolution, at best adding them post factum as an interesting but generally
unimportant perturbers. The real picture is likely to be much more complicated
as hinted by observations of ``transition discs''
\citep{AndrewsEtal11,OwenClarke12}, with massive planets not only taking the
mass from the disc but also giving it back when these objects are disrupted by
tides from the host star \citep{NayakshinLodato12,Nayakshin13a}. This cannot
possibly be added as a post factum perturbation to the disc evolution since
the number of gas fragments born in the outer disc per star may be significant
\citep[e.g.,][]{VB06} and the total mass in these fragments during the overall
disc lifetime be as much as $0.05 \msun$ or larger
\citep{ChaNayakshin11a}. This view joins up well with the developing paradigm
of episodic accretion of young protostars in which all stars consume a number
of massive clumps shipped into the star by disc migration from $\sim$ hundreds
of AU \citep{DunhamVorobyov12,AudardEtal14,VB15}.

\section{Conclusions}

Here we presented the third in the series paper on the population synthesis of
planet formation experiments in the framework of the Tidal Downsizing (TD)
model. A number of observed facts and correlations for exoplanet populations
are reproduced by our model. On this basis we believe that TD model is a
physically attractive alternative to CA in explaining many if not all of the
observed exoplanets. We presented predictions for future observations that may
distinguish this theory from CA. We hope that this work will stimulate more
theorists to contribute to the development of TD theory for planet formation.

\section{Acknowledgments}

Theoretical astrophysics research at the University of Leicester is supported
by a STFC grant. The authors acknowledge useful discussions with Richard
Alexander and Vardan Adibekyan.  This paper used the DiRAC Complexity system,
operated by the University of Leicester, which forms part of the STFC DiRAC
HPC Facility (www.dirac.ac.uk). This equipment is funded by a BIS National
E-Infrastructure capital grant ST/K000373/1 and DiRAC Operations grant
ST/K0003259/1. DiRAC is part of the UK National E-Infrastructure.

\end{document}